\documentclass[aps,pre,superscriptaddress,showkeys,amsmath,amssymb,twocolumn,nofootinbib]{revtex4-2} 
\usepackage{orcidlink}
\usepackage{bm}
\usepackage{microtype}
\usepackage[separate-uncertainty=true, multi-part-units=single]{siunitx}
\usepackage[version=4]{mhchem}
\usepackage{graphicx}
\graphicspath{{FigsL_Sanatan}}

\usepackage[svgnames, table, xcdraw]{xcolor}
\usepackage{hyperref}
\definecolor{customblue}{HTML}{2E3092}
\hypersetup{colorlinks, breaklinks, urlcolor=customblue, linkcolor=customblue, citecolor=customblue, anchorcolor=customblue} 

\begin{document}
\title{Identifying the signatures of residual activity in harmonically bound active Brownian dynamics}
\author{Sanatan Halder~\orcidlink{0009-0002-2457-0449}}
\email{sanatanh@iitk.ac.in}
\affiliation{Department of Physics, Indian Institute of Technology Kanpur, Kanpur -- 208016, India}
\author{Manas Khan~\orcidlink{0000-0001-6446-3205}}
\email{mkhan@iitk.ac.in}
\affiliation{Department of Physics, Indian Institute of Technology Kanpur, Kanpur -- 208016, India}

\begin{abstract}
	A confined self-propelled particle exhibits a range of intriguing dynamical phenomena dictated by the interplay between the intrinsic activity of the particle and the imposed confinement. This competition manifests as a crossover in the steady-state position distribution of a harmonically bound active Brownian particle (HBABP) from Boltzmann-like to bimodal, commonly recognized as the passive and active regimes, respectively, upon variations in activity and confinement strength. We present a comprehensive analysis of the resultant dynamics of an HBABP employing analytical calculations and numerical simulations, examining the variations in the position distribution, residual or resultant velocity, mean square displacement, power spectral density, and effective harmonic confinement at varying activities in the characteristic regimes across the crossover. These analyses provide a reliable identification of the signature of residual or remnant activity in ABP dynamics after being impeded by the harmonic confinement. Our results show that the resultant HBABP dynamics in the regime with a Boltzmann-like position distribution is dominated by residual activity, and the motion in the other regime, with a bimodal position distribution, is similar to that of a harmonically bound Brownian particle--devoid of residual activity--at a displaced position, where the activity is balanced by the restoring force field. 
	
\end{abstract}
\maketitle

\section{Introduction}\label{sec:intro}

The active dynamics of natural microswimmers, such as bacteria, protozoa, and spermatozoa, and their synthetic counterparts, self-propelled microparticles, remain an important topic of contemporary research because of their relevance in describing fundamental nonequilibrium processes and potential applications~\cite{Ramaswamy2010,Romanczuk2012, Marchetti2013, Cates2015, Bechinger2016, Fodor2018}. Self-propelled microparticles are conveniently modeled as active Brownian particles (ABPs), where the propulsion speed remains constant and the direction evolves with the orientational diffusion of the particles~\cite{Romanczuk2012, Bechinger2016, Fodor2018, basuActiveBrownianMotion2018}. Thus, the directional correlation decays over a characteristic timescale, called the persistence time.

An isolated ABP manifests effective passive-like behaviors at long times as the propulsion direction becomes random unless it interacts with an external potential~\cite{basuActiveBrownianMotion2018, tenhagenBrownianMotionSelfpropelled2011}. A confining potential impedes its active dynamics, resulting in a plethora of intricate dynamical phenomena, e.g., accumulation toward the periphery~\cite{Berke2008, elgetiWallAccumulationSelfpropelled2013,leeActiveParticlesConfinement2013}, self-induced polar ordering~\cite{bauerleFormationStableResponsive2020,lavergneGroupFormationCohesion2019,hennesSelfInducedPolarOrder2014}, and anomalous sedimentation~\cite{solonActiveBrownianParticles2015, Ginot2015}. Understanding the response of self-propelled particles under confinement remains a central question in active matter physics because they represent the interaction of natural and synthetic active matter with their environment, such as narrow confinements, crowded, porous, and viscoelastic media~\cite{filyDynamicsSelfpropelledParticles2014, BenIsaac2015, Bechinger2016, Ribeiro2018, narinderActiveParticlesGeometrically2019, Sprenger2022, Moore2023}. Harmonic confinement, which also approximately represents other potentials near a stable point, engenders intriguing nonequilibrium features in ABP dynamics, even in the steady state, by introducing an additional timescale, i.e., the equilibration time~\cite{tenhagenBrownianMotionSelfpropelled2011, basuActiveBrownianMotion2018,caraglioAnalyticSolutionActive2022,basuLongtimePositionDistribution2019}. Harmonically bound ABPs (HBABPs) have been extensively investigated in recent times, exploring the competition between confinement and activity using analytical~\cite{pototskyActiveBrownianParticles2012, dauchotDynamicsSelfPropelledParticle2019, chaudhuriActiveBrownianParticle2021, malakarSteadyStateActive2020, Santra2021, caraglioAnalyticSolutionActive2022, Baldovin2023, dasConfinedActiveBrownian2018}, numerical~\cite{pototskyActiveBrownianParticles2012, malakarSteadyStateActive2020, Santra2021, buttinoniActiveColloidsHarmonic2022}, and experimental~\cite{takatoriAcousticTrappingActive2016, schmidtNonequilibriumPropertiesActive2021, buttinoniActiveColloidsHarmonic2022} approaches.

Existing studies have mostly focused on the crossover in the steady-state position distribution from Boltzmann-like to bimodal as the harmonic confinement becomes stronger relative to the activity of the ABP. Considering the shape of the distributions, this crossover is commonly interpreted as passive to active or equilibrium-like to strongly nonequilibrium transition~\cite{pototskyActiveBrownianParticles2012, takatoriAcousticTrappingActive2016, basuLongtimePositionDistribution2019, malakarSteadyStateActive2020, chaudhuriActiveBrownianParticle2021, Santra2021, schmidtNonequilibriumPropertiesActive2021, buttinoniActiveColloidsHarmonic2022, Baldovin2023}. Reentrant behavior, in which the Boltzmann-like position distribution reappears under strong confinement, has also been predicted~\cite{chaudhuriActiveBrownianParticle2021,malakarSteadyStateActive2020}. However, whether the shape of the steady-state position distribution alone can provide a reliable signature of activity in the resultant dynamics of an HBABP or serve as a measure of how far it is from equilibrium remains questionable. A free ABP, being a strongly nonequilibrium system, also exhibits a Boltzmann-like position distribution in the leading order at times much longer than the persistence time~\cite{basuActiveBrownianMotion2018,Bechinger2016,Fodor2018, basuLongtimePositionDistribution2019}. Moreover, it appears inconsistent that ABPs manifest equilibrium-like or passive-like behavior under weak confinement while revealing nonequilibrium properties when the restoring force field becomes stronger. Therefore, a detailed dynamical analysis is essential for reliably identifying and measuring the signature of activity in the resultant HBABP dynamics.

In this study, we provide a comprehensive dynamical analysis of the resultant dynamics of an HBABP by employing analytical calculations and numerical simulations to reliably identify and measure the presence of activity therein. Alongside the variations in the position distributions, we analyze the resultant velocity components, mean square displacements (MSDs), effective harmonic confinements, and power spectral density (PSD) at varying activities and strengths of confinement to demonstrate that the shape of the position distribution is not a reliable signature of the residual activity, i.e., the remnant activity in the resultant dynamics of an HBABP. With appropriate identification of the manifestation of residual activity, we show that the regime dominated by residual activity exhibits a Boltzmann-like position distribution that broadens with activity, whereas a bimodal position distribution is observed in the residual activity-depleted regime. In a companion \textit{Letter}, we present experimental results validating this conclusion and demonstrate that the crossover from Boltzmann-like to bimodal steady-state position distribution is solely governed by the interplay between the characteristic timescales: the persistence time $\tau_{\mathrm{R}}$ and the equilibration time in the harmonic confinement $\tau_k$, i.e., by the ratio $\tau_{\mathrm{R}} / \tau_k$~\cite{Halder2025a}. 

We show that at $\tau_{\mathrm{R}} \ll \tau_k$ (regime - I), a strong presence of residual activity in the resultant HBABP dynamics is manifested by the broadening of the steady-state position distribution with increasing activity, large resultant velocity components that add up to the propulsion speed of the ABP, ballistic rise of MSDs at intermediate time scales, and a deviation of the PSD from the fluctuation-dissipation theorem prediction for a Brownian particle. In contrast, at $\tau_{\mathrm{R}} \gg \tau_k$ (regime - II), bimodal position distributions, where the peaks maintain their width but shift apart with increasing propulsion speed, resultant velocity components being similar to those in the case of the corresponding passive system, i.e., propulsion speed $V$ = 0, or a harmonically bound Brownian particle (HBBP), significantly smaller MSDs, effective harmonic confinement with unaltered strength but progressively displaced center with activity, and Lorentzian PSDs, which is a signature of HBBP dynamics, indicate that the resultant HBABP dynamics in this regime essentially reduces to the corresponding HBBP dynamics, devoid of residual activity, at a displaced position, where the activity is balanced by the restoring force field. In the intermediate regime ($\tau_{\mathrm{R}} = \tau_k$), these dynamical properties exhibit a smooth crossover.

The remainder of this article is organized as follows. Sec. \ref{sec:theory} provides a theoretical description of the HBABP system, outlines the Langevin dynamics simulation protocol, and details its experimental realization. Sec. \ref{sec:dyn_regs} identifies and analyzes the signature of activity in the resultant HBABP dynamics at A. $\tau_{\mathrm{R}} \ll \tau_k$ (regime - I), B. $\tau_{\mathrm{R}} \gg \tau_k$ (regime - II), and C. $\tau_{\mathrm{R}} = \tau_k$ (intermediate regime). Finally, Sec. \ref{sec:conclusions} summarizes our findings and discusses their implications. Detailed analytical derivations are provided in the appendix.

\section{System description}\label{sec:theory}

\subsection{Theoretical description}

\begin{figure}[ht]
	\centering
	\includegraphics[width=88mm]{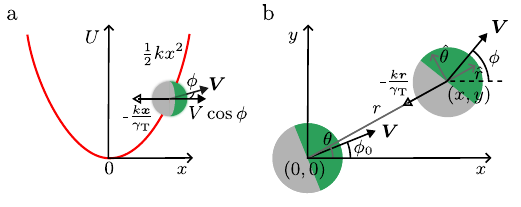}
	\caption{An active Brownian particle (ABP) in a harmonic well (HW). (a)~Schematic shows an ABP, depicted as an active Janus colloid (gray-green), in an HW with potential $U = \frac{1}{2} k x^2$ (orange curve). The corresponding restoring velocity ($ = -kx/\gamma_{\mathrm{T}}$) is shown as an open arrow. The Janus particle undergoes active propulsion with velocity $\bm{V}$, which has a constant speed and points from the coated side (gray) to the bare side (green), making an angle $\phi$ with the $x$-axis in the $x-y$ plane. $\bm{V}$ and its $x$-component $\bm{V} \cos{\phi}$ are represented by the solid arrows. (b)~Active propulsion velocity $\bm{V}$ (solid arrows) and the restoring velocity ($ = - k \bm{r} / \gamma_{\mathrm{T}}$, open arrow) are shown in the $x-y$ plane for the initial ($t = 0$) and representative position and orientation at time $t$. Starting from the origin (0, 0) with orientation $\phi_0$ at $t = 0$, the position and orientation of the Janus particle evolve to ($x, y$) and $\phi$, respectively, at time $t$. The particle position in plane-polar coordinates (gray) is given by ($r, \theta$).}%
	\label{fig:HBABP}%
\end{figure}

An ABP experiences self-propulsion at a constant speed $V$ along a body-fixed direction that reorients through the orientational diffusion of the particle with coefficient $D_{\mathrm{R}}$, along with thermally exited Brownian motion. Therefore, the persistence time of the active dynamics, over which the correlation of the propulsion direction decays, is given by $\tau_{\mathrm{R}} = 1/D_{\mathrm{R}}$ ~\cite{howseSelfMotileColloidalParticles2007,jiangActiveMotionJanus2010, tenhagenBrownianMotionSelfpropelled2011, Bechinger2016, basuActiveBrownianMotion2018, Halder2026}. The relative strength of the activity is characterized by the P\'{e}clet number, $\mathrm{Pe} = V / \sqrt{D_{\mathrm{R}} D_{\mathrm{T}}}$, where  $D_{\mathrm{T}}$ is the translational diffusion coefficient.

The confinement by a 2D isotropic harmonic well (HW) $U(r) = (1/2) k r^2$ with force constant $k$ (Fig.~\ref{fig:HBABP}), corresponding to an equilibration time of $\tau_k = \gamma_{\mathrm{T}} / k$, where $\gamma_{\mathrm{T}}$ is the translational Stokes drag coefficient, impedes the active dynamics of the ABP. The resultant translational ($x(t)$, $y(t)$) and orientational ($\phi(t)$) dynamics of the HBABP are described by Langevin equations ~\cite{tenhagenBrownianMotionSelfpropelled2011}:
\begin{align}
	\begin{split}
		\dot{x}    & = v_{\mathrm{B}}^x - x / \tau_k + V \cos{\phi }, \\
	    \dot{y}    & = v_{\mathrm{B}}^y - y / \tau_k + V \sin{\phi}, \ \; \text{and} \\
		\dot{\phi} & =  v_{\mathrm{B}}^{\phi},
		\label{eq:LE}
	\end{split}
\end{align}
\noindent
where $v^i_{\mathrm{B}}(t)$ ($i = x, y, \phi$) are the Brownian velocities with $\langle v^i_{\mathrm{B}}(t) \rangle = 0$ and $\langle v^i_{\mathrm{B}}(t_1) v^j_{\mathrm{B}}(t_2) \rangle = 2 D_i \delta_{ij} \delta(t_1 - t_2)$, with $D_x = D_y = D_{\mathrm{T}}$ and $D_\phi = D_{\mathrm{R}}$. Here, the orientation of the ABP $\phi(t)$ is independent its translational dynamics and is governed by free Brownian diffusion, given by the time-dependent probability distribution $P(\phi, \phi_0; t) = (4 \pi D_{\mathrm{R}} t)^{-1/2} \exp\left[ -(\phi - \phi_0)^2 / 4 D_{\mathrm{R}} t \right]$, where $\phi_0 = \phi(0)$ \cite{tenhagenBrownianMotionSelfpropelled2011}.

The translational equation for $x(t)$ (Eq.~\ref{eq:LE}) can be solved to obtain~\cite{tenhagenBrownianMotionSelfpropelled2011, doiSoftMatterPhysics2013}
\begin{align}
	x(t) & = e^{-t/\tau_k} \left[ x_0 + \int_{0}^{t} e^{t^{\prime}/\tau_k} \left\lbrace v^x_{\mathrm{B}}(t^{\prime}) + V\cos\phi(t^{\prime}) \right\rbrace dt^{\prime}\right] \nonumber \\
	     & = e^{-t/\tau_k} \left[ x_0 + \int_{0}^{t} e^{t^{\prime}/\tau_k} A^x (t^{\prime})\, dt^{\prime}\right],
	\label{eq:x}
\end{align}
\noindent where $x_0 = x(0)$, and $A^x(t) = v^x_{\mathrm{B}}(t) + V \cos\phi(t)$ comprises both thermal and active velocity noises along $x$. The solution for $y(t)$ is obtained in a similar fashion, with $A^y(t) = v^y_{\mathrm{B}}(t) + V \sin\phi(t)$. Since Brownian velocity noises $v^i_{\mathrm{B}}$ are uncorrelated with $\phi(t)$, the cross-terms vanish. Consequently, the mean values and autocorrelations of $A^x$ and $A^y$ can be evaluated analytically (Appendix~\ref{appendix:a}). The closed-form solutions for the position distributions $P(x ; t)$ and $P(y ; t)$ can then be obtained in the two limiting regimes $\tau_{\mathrm{R}} \ll \tau_k$ and $\tau_{\mathrm{R}} \gg \tau_k$, as shown in Appendix~\ref{appendix:b}.

At $\tau_{\mathrm{R}} \ll \tau_k$, $P (x ; t)$ and $P (y ; t)$ take the same form, which is given by,
\begin{equation}
	P (x, x_0; t) = \frac{1}{\sqrt{2 \pi \sigma_{1, t}^2}} \exp \left[ - \frac{\left(x - x_0 e^{-t/\tau_k} \right)^2}{2 \sigma_{1, t}^2}\right],
 \label{eq:Pxt1}
\end{equation}
\noindent where $\sigma_{1, t}^2 = \frac{k_{\mathrm{B}}T}{k} \left(1 + \frac{\mathrm{Pe}^2}{2}\right) \left( 1 - e^{- 2t / \tau_k} \right)$. In the steady-state, i.e., at $t \gg \tau_k$, $\sigma_{1, t}^2$ becomes  $\sigma_1^2 = \frac{ k_{\mathrm{B}} T}{k}\left(1 + \frac{\mathrm{Pe}^2}{2}\right) $, and the position distributions reduce to,
\begin{equation}
	P_{\mathrm{ss}} (x) = \frac{1}{\sqrt{2 \pi \sigma_{1}^2}} \exp \left[ - \frac{x^2}{2 \sigma_{1}^2}\right].
	\label{eq:PxSS1}
\end{equation}
Thus, the 1D position distributions become Gaussian at steady-state, and the 2D counterpart remains Boltzmann-like,
\begin{equation}
	P_{\mathrm{ss}} (r) = P_{\mathrm{ss}} (x) P_{\mathrm{ss}} (y) = \frac{1}{2 \pi \sigma_{1}^2} \exp \left[ - \frac{r^2}{2 \sigma_{1}^2}\right].
	\label{eq:PrSS1}
\end{equation}

However, at $\tau_{\mathrm{R}} \gg \tau_k$, the forms of $P (x)$ and $P (y)$ depend on the initial propulsion direction $\phi_0$. Considering $\phi_0 = 0$, i.e., the initial propulsion direction to be along $\hat{x}$, $P (x)$ and $P (y)$ evolve differently as,
\begin{equation}
\begin{split}
& P (x, x_0; t) = \frac{1}{\sqrt{2 \pi \sigma_{2, t}^2}} \times \\
&  \exp \left[ - \frac{\left(x - x_0 e^{-t/\tau_k} - V \tau_k \left( 1 - e^{-t/\tau_k}\right) \right)^2}{2 \sigma_{2, t}^2}\right] 
\end{split} 
\label{eq:Pxt2}
\end{equation}
\noindent and
\begin{equation}
P (y, y_0; t) = \frac{1}{\sqrt{2 \pi \sigma_{2, t}^2}} \times \exp \left[ - \frac{\left(y - y_0 e^{-t/\tau_k}\right) ^2}{2 \sigma_{2, t}^2} \right],
\label{eq:Pyt2}
\end{equation}
where $\sigma_{2, t}^2 = \sigma_{\mathrm{HBBP}, t}^2 = D_{\mathrm{T}} \tau_k \left( 1 - e^{- 2t / \tau_k} \right)$. Both $P (x)$ and $P (y)$ take the form of that of a HBBP at $ t \gg \tau_k$ (and $t \ll \tau_{\mathrm{R}}$), albeit $P (x)$ being shifted by $V \tau_k$, and are given by,
\begin{equation}
	P(x) = \frac{1}{\sqrt{2 \pi \sigma_{2}^2}} \exp \left[ - \frac{ \left( x - V \tau_k \right) ^2}{2 \sigma_{2}^2}\right], 
	\label{eq:PxSS2}
\end{equation}
\noindent and
\begin{equation}
P(y) = \frac{1}{\sqrt{2 \pi \sigma_{2}^2}} \exp \left[ - \frac{ y^2}{2 \sigma_{2}^2}\right],
\label{eq:PySS2}
\end{equation}
\noindent where the time-independent variance $\sigma_2^2 = \sigma_{\mathrm{HBBP}}^2 = k_{\mathrm{B}} T / k$. At $t \gg \tau_{\mathrm{R}}$, i.e., at steady-state, the anisotropy in $P (x)$ and $P (y)$ go away with full orientational diffusion of the ABP, and the 2D position distribution becomes
\begin{equation}
	P_{\mathrm{ss}}(r) = \frac{1}{2 \pi \sigma_{2}^2} \exp \left[ - \frac{ \left( r - V \tau_k \right) ^2}{2 \sigma_{2}^2}\right].
	\label{eq:PrSS2}
\end{equation}
The position distribution covers an annular region and appears bimodal when projected onto 1D.

The resultant velocity, given by the Cartesian components $\dot{x}$ and $\dot{y}$ in Eq.~\ref{eq:LE} describes the residual velocity, $v_{\mathrm{res}}$ (i.e., $v_{\mathrm{res}}^{x} = \dot{x}$ and $v_{\mathrm{res}}^{y} = \dot{y}$) of the ABP after it encounters the restoring force field. Therefore, $v_{\mathrm{res}}$ provides a convenient quantitative measure of the active dynamics that survives the harmonic confinement. Exploiting the circular symmetry of the HW, we analyzed the radial and azimuthal components of $v_{\mathrm{res}}$, i.e., $v_{\mathrm{res}}^{r}$ and $v_{\mathrm{res}}^{\theta}$, respectively. They are given by the Langevin equations (Eq.~\ref{eq:LE}) in plane-polar coordinates as,
\begin{equation}
	v_{\mathrm{res}}^{r} = V^{r} - r / \tau_k + v_{\mathrm{B}}^{r} \quad \text{and} \quad v_{\mathrm{res}}^{\theta} = V^{\theta} + v_{\mathrm{B}}^{\theta}.
	\label{eq:ResVel}
\end{equation}
The radial restoring force field only affects $v_{\mathrm{res}}^{r}$, whereas $v_{\mathrm{res}}^{\theta}$ does not experience the harmonic confinement, resulting in free azimuthal dynamics.

The closed-form solution for the ensemble-averaged 2D MSD of an HBABP can be derived from the velocity autocorrelations as (Appendix~\ref{appendix:c}),
\begin{align}
	\left\langle \Delta r^{2} (\tau) \right\rangle & = \frac{4 k_{\mathrm{B}} T}{k}\left[ 1 - e^{- \tau / \tau_{k}} \right] \nonumber \\
	& \quad +  \frac{2 V^2 \tau_{k}^2 \tau_{\mathrm{R}}}{\tau_{\mathrm{R}} + \tau_{k}}  \left[ 1 - \frac{\tau_{\mathrm{R}} e^{- \tau / \tau_{\mathrm{R}}} - \tau_{k} e^{- \tau / \tau_{k}}}{\tau_{\mathrm{R}} - \tau_{k}} \right],
	\label{eq:MSDGen}
\end{align}
\noindent where the first term, which is the same as the MSD of an HBBP, corresponds to the contribution of passive dynamics, and the second term describes the effect of the activity. Eq.~\ref{eq:MSDGen} agrees with the existing results~\cite{jiangActiveMotionJanus2010} and reduces to simplified forms in the two limiting regimes, $\tau_{\mathrm{R}} \ll \tau_k$ and $\tau_{\mathrm{R}} \gg \tau_k$.

At $\tau_{\mathrm{R}} \ll \tau_k$, it takes the form
\begin{equation}
	\begin{split}
		\left\langle \Delta r^{2} (\tau) \right\rangle  = & \left( \frac{4 k_{\mathrm{B}} T}{k} + 2 V^2 \tau_{k} \tau_{\mathrm{R}}\right) \left[ 1 - e^{- \tau / \tau_{k}}\right] \\
		& \qquad + 2 V^2 \tau_{\mathrm{R}}^2 e^{- \tau / \tau_{\mathrm{R}}}.
	\end{split}
	\label{eq:MSD1}
\end{equation}
\noindent At long time, the MSD reaches a plateau, which is given by,
\begin{equation}
	\left\langle \Delta r^{2} \right\rangle_{\mathrm{ss}} = \frac{4 k_{\mathrm{B}} T}{k} + 2 V^2 \tau_{k} \tau_{\mathrm{R}}.
	\label{eq:MSD1ss}
\end{equation}

At the other limit, i.e., at $\tau_{\mathrm{R}} \gg \tau_k$, the MSD becomes,
\begin{equation}
	\left\langle \Delta r^{2} (\tau) \right\rangle = \frac{4 k_{\mathrm{B}} T}{k}\left[1 - e^{-\tau /\tau_k}\right] + 2V^2 \tau_k^2 \left[1 - e^{-\tau /\tau_{\mathrm{R}}}\right],
	\label{eq:MSD2}
\end{equation}
\noindent and the corresponding long-time plateau value is,
\begin{equation}
	\left\langle \Delta r^{2} \right\rangle_{\mathrm{ss}} = \frac{4 k_{\mathrm{B}} T}{k} + 2V^2 \tau_k^2.
	\label{eq:MSD2ss}
\end{equation}

By comparing the form of the steady-state position distributions and MSDs with those of an HBBP, effective harmonic confinements can be defined for the two regimes, $\tau_{\mathrm{R}} \ll \tau_k$ and $\tau_{\mathrm{R}} \gg \tau_k$. The strength of the effective harmonic confinement $k_{\mathrm{eff}}$ and the radial position of its center $r_{\mathrm{c}}$ are given by,
\begin{align}
		k_{\mathrm{eff}} & = \frac{k}{ 1 + \mathrm{Pe}^2 / 2},  &  r_{\mathrm{c}} & = 0 &  \text{for}    \ \tau_{\mathrm{R}} \ll \tau_{k}, 
		\label{eq:k_eff1}\\ 
		k_{\mathrm{eff}}  &  = k,  & r_{\mathrm{c}} & = V \tau_k & \text{for} \ \tau_{\mathrm{R}} \gg \tau_{k}.
		\label{eq:k_eff2}
\end{align}

The single-sided power spectral densities, $\mathrm{PSD}_x = |X(f)|^2$ and $\mathrm{PSD}_y = |Y(f)|^2$, where $X(f)$ and $Y(f)$ are the Fourier transforms of $x(t)$ and $y(t)$, characterize the frequency response of the HBABP dynamics. Following the fluctuation-dissipation theorem (FDT), the PSDs of an HBBP, i.e., of an HBABP with $V$ = 0, are given by Lorentzian,
\begin{equation}
	\mathrm{PSD}_x = \mathrm{PSD}_y = \frac{k_{\mathrm{B}}T}{\pi^2 \gamma_{\mathrm{T}} \left( f^2 + f_{0}^2 \right)} = \frac{L_{0}^2}{f^2 + f_{0}^2},
	\label{eq:PSD_HBBP} 
\end{equation}
\noindent where $f_0 = (2 \pi \tau_k)^{-1}$ is the corner frequency~\cite{Halder2024}. Thus, deviations of the HBABP PSDs from this Lorentzian (Eq.~\ref{eq:PSD_HBBP}) serve as a frequency-dependent signature of residual activity.

\subsection{Langevin dynamics simulation}\label{sec:sim}

We numerically simulated Eq.~\ref{eq:LE} with the desired values of $V$, $\tau_{\mathrm{R}}$, and $\tau_{k}$, where $\tau_{\mathrm{R}}$ determined $D_{\mathrm{R}}$ and consequently $D_{\mathrm{T}}$, with a consistent particle size and viscosity ($\eta$) value of that of water. The orientational Brownian displacement $\Delta \phi$ in time-step $\Delta t$ was considered a normal random variable with a mean of zero and a standard deviation of $\sqrt{2 D_{\mathrm{R}} \Delta t}$. This provided the orientational time-series $\phi (t_i) = \phi (t_{i-1}) + \Delta \phi (\Delta t)$, where $t_i = t_{i-1} + \Delta t$. The translational dynamics were then simulated by adding propulsion components to the corresponding HBBP dynamics along both $x$ and $y$ at each time-step, as $x_{\mathrm{HBABP}} (t_i) = x_{\mathrm{HBBP}} (t_i) + V \cos(\phi (t_{i-1})) \Delta t$. The HBBP dynamics, i.e., $x_{\mathrm{HBBP}} (t_i)$ and $y_{\mathrm{HBBP}} (t_i)$, with given $D_{\mathrm{T}}$ and $\tau_{k}$, were conveniently obtained by employing Green's function:
\begin{equation*}
	G (x (t_i), x (t_{i-1}); \Delta t) =  \frac{1}{\left[ 2 \pi B\right] ^{1/2}} \exp\left( -\frac{(x (t_i)- A)^2}{2B}\right),
\end{equation*}
which represents a Gaussian probability with time-dependent mean $A (\Delta t) = x (t_{i-1}) \exp(- \Delta t / \tau_k)$ and variance $B (\Delta t) = \frac{k_{\mathrm{B}}T}{k} (1-\exp(-2 \Delta t/ \tau_k))$~\cite{Chandrasekhar1943, Khan2014a, Khan2014}. Simulated HBABP trajectories were used for further analyses.

We simulated HBBAP dynamics at varying $V$ with values of 0, 1, 2, 5, and \SI{10}{\micro\meter\per\second} for three cases: $\tau_{\mathrm{R}}$ (= \SI{1}{\second}) $\ll$ $\tau_{k}$ (= \SI{100}{\second}), $\tau_{\mathrm{R}}$ (= \SI{100}{\second}) $\gg$ $\tau_{k}$ (= \SI{1}{\second}), and at $\tau_{\mathrm{R}} = \tau_{k}$ = \SI{10}{\second}. The Stokes radius ($a$) of the ABP, and consequently, its translational diffusion coefficient ($D_{\mathrm{T}}$) and the stiffness of the harmonic potential ($k$)  were set accordingly, keeping the medium viscosity ($\eta$ = \SI{1}{\m\Pa.\s}) and temperature ($T$ = \SI{300}{\K}) unchanged.

Notably, choosing an appropriately short time-step is crucial for simulating ABP or HBABP dynamics because of the persistence in the propulsion direction. By simulating ABP trajectories with progressively shorter $\Delta t$ and analytically calculating the rms value of the deviation in the active dynamics during $\Delta t$, we concluded that the deviation in the trajectory would become negligible ($\lesssim$ 0.5\%) with a further shortening of $\Delta t$ beyond $\Delta t \approx 10^{-2} \times \tau_{\mathrm{R}}$. Moreover, to capture the details of the equilibration dynamics of an ABP in the HW, we used $\Delta t$ that satisfied $\Delta t \lesssim (10^{-2} \times \tau_{\mathrm{R}}, 10^{-2} \times \tau_{k})$ in all our simulations.

While calculating the instantaneous values of $v_{\mathrm{res}}^r (t) = \dot{r}(t)$ and $v_{\mathrm{res}}^{\theta} (t) = r(t) \dot{\theta}(t)$ from the simulated trajectories, we considered the average values of $r(t)$ and $\theta(t)$ over \SI{0.1}{\s} to eliminate the higher-frequency fluctuations. The averaging time window was chosen to be \SI{0.1}{\s}, as it is at least ten times shorter than the characteristic timescales ($\tau_{\mathrm{R}}$, $\tau_{k}$) used in the simulations, ensuring that $v_{\mathrm{res}}$ retains all the relevant and crucial dynamical information. Furthermore, to compute the radial position distribution $P (r)$ and PSD in the regime where the HBABP dynamics are confined in an annular region, a part of the trajectories in narrow strips along the radial directions were considered.

\begin{figure*}[ht]
	\centering
	\includegraphics[width=175mm]{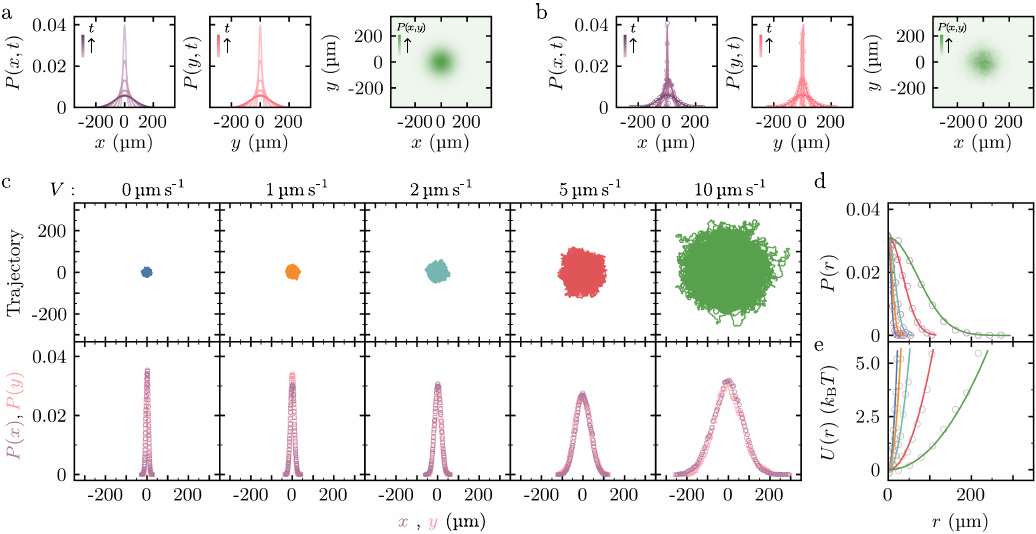}
	\caption{Position distributions in regime - I, where $\tau_{\mathrm{R}} \ll \tau_k$  ($\tau_{\mathrm{R}}/\tau_{k} = 0.01$; $\tau_{\mathrm{R}} =$ \SI{1}{\s}, $\tau_{k} =$ \SI{100}{\s}). (a)~Analytically predicted time evolution of $P(x, t)$ (left) and $P(y, t)$ (middle) for $V =$ \SI{10}{\micro\meter\per\second}, displayed as color-coded curves at successive times $t$. The 2D steady-state distribution following Eq.~\ref{eq:PrSS1} is shown as a density plot (right).  (b)~Numerically simulated position distributions at increasing times are shown in the same way. (c)~Five simulated trajectories of \SI{E5}{\s} duration, generated at \SI{E3}{\Hz}, with propulsion speeds $V$ varying from $V$ = 0 (HBBP) to \SI{10}{\micro\meter\per\second}, are displayed in blue, orange, cyan, red, and green, respectively (top panel). The corresponding steady-state position distributions $P(x)$ (circles) and $P(y)$ (squares) are shown in the bottom panel. (d)~ Radial position distributions $P(r)$ (circles), obtained from the simulated trajectories, are exhibited with Gaussian curve fittings (color-coded solid lines), following Eq.~\ref{eq:PrSS1} (solid lines). (e)~Effective confining potentials $U(r)$, computed from the respective $P(r)$ by Boltzmann inversion (open circles), exhibit excellent fitting to $\frac{1}{2} k_{\mathrm{eff}}\,(r - r_{\mathrm{c}})^2$ (color-coded solid lines), where $r_{\mathrm{c}}$ remains zero and $k_{\mathrm{eff}}$ decreases with increasing $V$, in accordance with Eq.~\ref{eq:k_eff1}.}%
	\label{fig:PosDist1}%
\end{figure*}

\subsection{Experimental realization}\label{sec:expt}

ABPs were experimentally realized using Pt-silica Janus microspheres that exhibited diffusiophoretic self-propulsion in a $\mathrm{H_2O_2}$ suspension and thermophoretic active motion under laser exposure. The harmonic well was set up using an optical trap. Thus, a Janus particle in the optical trap provided experimental realization of an HBABP, where the strength of the confinement ($k$) and thermophoretic activity ($V$) were regulated by tuning the laser power, and self-propulsion speed ($V$) was supplemented by adding $\mathrm{H_2O_2}$ at an appropriate concentration~\cite{Halder2025a, Halder2026}. The symmetry of the harmonic confinement of a Janus particle by the optical trap was validated by the position distribution, and the activity and confinement parameters $V$, $\tau_{\mathrm{R}}$, and $\tau_k$ were determined by fitting the experimental MSDs to Eq.~\ref{eq:MSDGen} \cite{Halder2025a, Halder2026}. All experimental observations and results are discussed in detail in the companion Letter~\cite{Halder2025a}.

\section{Signatures of activity}\label{sec:dyn_regs}

We studied the manifestation of the remnant activity of an HBABP in different dynamical regimes with disparate steady state position distributions – Boltzmann-like position distributions that are observed in the regime $\tau_{\mathrm{R}} \ll \tau_k$ (Eq.~\ref{eq:PxSS1}, \ref{eq:PrSS1}), and bimodal position distributions emerging at $\tau_{\mathrm{R}} \gg \tau_k$ (Eq.~\ref{eq:PxSS2}, \ref{eq:PrSS2}), including an intermediate regime where $\tau_{\mathrm{R}} = \tau_k$. By analyzing the shape of the position distribution, resultant or residual velocity, MSDs, effective harmonic confinement, and PSDs, we examined the signatures of activity, i.e., the nonequilibrium indicators in the resultant dynamics of the HBABP in the above-mentioned three regimes. In this section, we discuss the analytical predictions and present the Langevin dynamics simulation results from the three representative regimes: $\tau_{\mathrm{R}} / \tau_k$ = 0.01 ($\tau_{\mathrm{R}}$ = \SI{1}{\s}, $\tau_k$ = \SI{100}{\s}), $\tau_{\mathrm{R}} / \tau_k$ = 100 ($\tau_{\mathrm{R}}$ = \SI{100}{\s}, $\tau_k$ = \SI{1}{\s}), and $\tau_{\mathrm{R}} / \tau_k$ = 1 ($\tau_{\mathrm{R}} = \tau_k$ = \SI{10}{\s}), where the propulsion speed is systematically varied from $V$ = 0 (corresponding to the passive counterpart, i.e., HBBP) to $V$ = \SI{10}{\micro\meter / \s}, showing their comprehensive agreement with each other. Notably, the corresponding experimental results are presented in the companion letter~\cite{Halder2025a}.

\subsection{Regime - I: $\tau_{\mathrm{R}} \ll \tau_k$}\label{sec:reg1}

This characteristic regime of an HBBAP with a Boltzmann-like position distribution in the steady state has been studied extensively and is commonly recognized as passive or equilibrium-like, considering the shape of the distribution~\cite{pototskyActiveBrownianParticles2012, takatoriAcousticTrappingActive2016, basuLongtimePositionDistribution2019, malakarSteadyStateActive2020, chaudhuriActiveBrownianParticle2021, Santra2021, schmidtNonequilibriumPropertiesActive2021, buttinoniActiveColloidsHarmonic2022, Baldovin2023}. Here, we present a detailed analysis of the resultant dynamics of the HBABP in this regime to reveal strong nonequilibrium signatures manifested by the presence of large residual activity.

\subsubsection{Position distribution}\label{sec:pos1}

When $\tau_{\mathrm{R}} \ll \tau_k$, $P (x; t)$ and $P (y; t)$ have the same form and widen symmetrically with time, as given by Eq.~\ref{eq:Pxt1} and shown in Fig.~\ref{fig:PosDist1}(a), until they reach a steady-state. At steady-state, the 1D and 2D position distributions become Gaussian (Eq.~\ref{eq:PxSS1}) and Boltzmann-like (Eq.~\ref{eq:PrSS1}), respectively, with variance $\sigma_1^2 = \frac{ k_{\mathrm{B}} T}{k}\left(1 + \frac{\mathrm{Pe}^2}{2}\right) $, which increases monotonically with activity $\mathrm{Pe}$. The steady-state position distribution widens, i.e., the spread of the long-time trajectory increases monotonically with the propulsion speed $V$, as shown by the solid lines in Fig.~\ref{fig:PosDist1}(d). This signifies the presence of active motion in the resultant dynamics, although the apparent Boltzmann-like shape of the position distribution is reminiscent of equilibrium or passive (i.e., HBBP) dynamics.

The numerically simulated time evolutions of $P (x; t)$ and $P (y; t)$, as shown in Fig.~\ref{fig:PosDist1}(b), agree excellently with the analytical predictions. The trajectories frequently cross the center as their spreads increase over time (Supp.~Video~~\href{https://drive.google.com/file/d/1RNH7BM5JcLSk9Q3rXtX4Jz0FPxZIqfZb/view?usp=drive_link}{1}) before reaching the steady-state. Typical steady-state trajectories with the corresponding $P(x)$ and $P(y)$ at five different propulsion speeds are exhibited in Fig.~\ref{fig:PosDist1}(c). The broadening of the bound trajectories and position distributions are also in accordance with the analytical predictions, corroborating the effect of activity on the steady-state position distributions. Moreover, the radial distributions $P (r)$ at varied $V$ match perfectly with the corresponding analytical forms (Fig.~\ref{fig:PosDist1}(d)).

These results, which show that the resultant HBABP dynamics exhibit a strong signature of activity in this regime despite having Boltzmann-like position distributions, are also corroborated by our experimental observations of optically trapped active Janus colloids~\cite{Halder2026, Halder2025a}.

\subsubsection{Residual velocity}\label{sec:resv1}

\begin{figure}[ht]
	\centering
	\includegraphics[width=85mm]{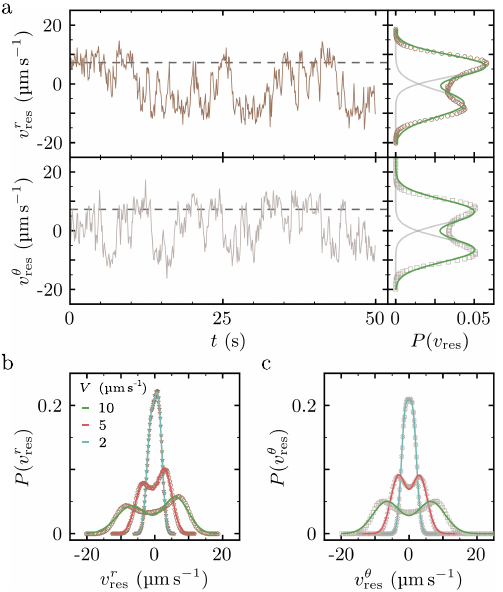}
	\caption{Residual, i.e., resultant velocity distributions for $\tau_{\mathrm{R}} \ll \tau_k$ (($\tau_{\mathrm{R}}/\tau_{k} = 0.01$; $\tau_{\mathrm{R}} =$ \SI{1}{\s}, $\tau_{k} =$ \SI{100}{\s})). (a)~Short segments (\SI{50}{\s}) of the time series (left) and probability distributions (right) of the radial component $v_{\mathrm{res}}^{r}$ (top) and azimuthal component $v_{\mathrm{res}}^{\theta}$ (bottom), calculated from the simulated trajectory with $V = \SI{10}{\micro\meter\per\second}$. The dashed horizontal lines in the time series represent the respective rms values. Both $P(v_{\mathrm{res}}^{r})$ and $P(v_{\mathrm{res}}^{\theta})$ exhibit bimodal shapes that fit well with the sum of two symmetrically positioned Gaussians (solid lines). (b, c)~Probability distributions $P(v_{\mathrm{res}}^{r})$ and $P(v_{\mathrm{res}}^{\theta})$ for three different propulsion speeds, $V =$ 2, 5, and \SI{10}{\micro\meter\per\second}. The bimodal shape of the distributions becomes progressively more apparent with increasing $V$, as the peak separation grows with the propulsion speed.}%
	\label{fig:ResVel1}%
\end{figure}

We obtained the components of the resultant or residual velocity $v_{\mathrm{res}}^r (t)$ and $v_{\mathrm{res}}^{\theta} (t)$ of the HBABP from the simulated trajectories at various propulsion speeds $V$. In this regime, both $v_{\mathrm{res}}^r$ and $v_{\mathrm{res}}^{\theta}$ exhibit wide fluctuations with bimodal distributions consisting of two symmetrically placed Gaussians at the corresponding rms values, as shown in Fig.~\ref{fig:ResVel1}(a). Notably, the mean-square values of the residual velocity components add up to the square of the propulsion speed, i.e., $\left\langle \left( v_{\mathrm{res}}^{r} \right) ^2\right\rangle + \left\langle \left( v_{\mathrm{res}}^{\theta} \right) ^2\right\rangle \approx V^2$, revealing that the ABP retains nearly all of its propulsion speed even under harmonic confinement. This is further corroborated by the fact that the separation between the peaks of the bimodal distributions of $v_{\mathrm{res}}^r$ and $v_{\mathrm{res}}^{\theta}$ becomes narrower with decreasing $V$, and the distributions eventually become flat-topped Gaussians (Figs.~\ref{fig:ResVel1}(b) and (c)). 

The $v_{\mathrm{res}}^r$ and $v_{\mathrm{res}}^{\theta}$ obtained from optically trapped active Janus colloids also exhibit flat-topped Gaussian distributions in this regime, and the rms values of $v_{\mathrm{res}}$ exceed the corresponding propulsion speeds by the Brownian velocity contribution, i.e., $v_{\mathrm{HBBP}}$~\cite{Halder2025a}. At $\tau_{\mathrm{R}} \ll \tau_{k}$, the restoring force does not lead to a sustained reduction in $v_{\mathrm{res}}^r$ because the active velocity rapidly changes its direction owing to a shorter $\tau_{\mathrm{R}}$. This rapid randomization of the propulsion direction, much before the ABP equilibrates in the harmonic well, results in Gaussian position distributions and prevents a steady balance between radial propulsion and the restoring force. Therefore, the activity remains largely undiminished and governs the resultant HBABP dynamics in this regime.

\subsubsection{Mean Square Displacement}\label{sec:msd1}

\begin{figure}[ht]
	\centering
	\includegraphics[width=72mm]{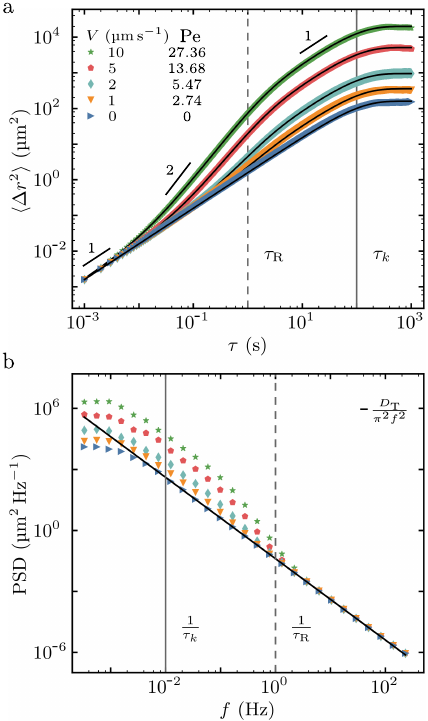}
	\caption{Mean square displacement (MSD) and power spectral density (PSD) in regime - I ($\tau_{\mathrm{R}}/\tau_{k} = 0.01$; $\tau_{\mathrm{R}} =$ \SI{1}{\s}, $\tau_{k} =$ \SI{100}{\s}). (a)~Computed MSDs, $\langle \Delta r^2 \rangle$, for five propulsion speeds ($V$), varying from 0 (HBBP) to \SI{10}{\micro\meter\per\second}, are shown with color-coded symbols consistent with the corresponding trajectories in Fig.~\ref{fig:PosDist1}c. The $V$ values and respective P\'{e}clet numbers ($\mathrm{Pe}$) are specified in the inset. Fitting to the analytical solution (Eq.~\ref{eq:MSDGen}) are shown with solid black lines. The characteristic timescales $\tau_{\mathrm{R}}$ and $\tau_{k}$ are marked with dashed and solid vertical lines, respectively, and the apparent slopes of 1 and 2 are indicated by short black straight lines as visual guides. 	(b)~PSDs, computed for these five propulsion speeds using the Welch method, are compared with the fluctuation-dissipation theorem (FDT) prediction for a free Brownian particle, given by $D_{\mathrm{T}} / \pi^2 f^2$ (black straight line). The vertical lines indicate the characteristic frequencies $1/\tau_{k}$ (solid) and $1/\tau_{\mathrm{R}}$ (dashed).}%
	\label{fig:MSDPSD1}%
\end{figure}

The MSD for this regime is expressed by Eq. \ref{eq:MSD1}. The MSD varies linearly with $\tau$ at very short times ($ \tau \ll \tau_{\mathrm{R}}$), where Brownian diffusion dominates, and again at $\tau_{\mathrm{R}} < \tau < \tau_k$, before eventually saturating to a plateau value, as given by Eq. \ref{eq:MSD1ss}, at $\tau > \tau_k$. At intermediate time-lags, the last term dominates, providing a ballistic $V^2 \tau^2$ dependence that directly manifests the presence of active propulsion in the resultant HBABP dynamics.  

The simulated MSDs at various $V$ values, along with their fit to Eq. \ref{eq:MSDGen}, are shown in Fig.~\ref{fig:MSDPSD1}(a). They exhibit the clear emergence of a ballistic $\tau^2$ regime at $\tau < \tau_{\mathrm{R}}$ for higher $V$ values. All MSDs exhibit linear growth at short time-lag $\tau \ll \tau_{\mathrm{R}}$ and for $\tau_{\mathrm{R}} < \tau < \tau_k$, reaching a plateau at $\tau > \tau_k$ (Fig.~\ref{fig:MSDPSD1}(a)). Signatures of active dynamics are also observed in the MSDs of optically trapped active Janus colloids in this regime ($ \tau_{\mathrm{R}} / \tau_k \ll$ 1)~\cite{Halder2025a}.

\subsubsection{Effective harmonic confinement}\label{sec:keff1}

The effective harmonic confinement $k_{\mathrm{eff}}$ of the HBABP in this regime is given by Eq. \ref{eq:k_eff1}. Here, a monotonic decrease in $k_{\mathrm{eff}}$ with $\mathrm{Pe}$ is a clear signature of the activity-governed resultant dynamics. We obtained the effective harmonic potential $U (r)$ through the Boltzmann inversion of the simulated steady-state position distribution $P(r)$ at various $V$ values, which are in complete agreement with our analytical description of $k_{\mathrm{eff}}$, as shown in Fig.~\ref{fig:PosDist1}(e). Furthermore, the variation of $k_{\mathrm{eff}}$, obtained from the fitting of the corresponding $U (r)$, with $V$ match excellently with our analytical prediction (Fig. \ref{fig:k_eff}). A decrease in $k_{\mathrm{eff}}$ with increasing activity, as described by Eq.~\ref{eq:k_eff1} has also been experimentally verified~\cite{Halder2026}.

\subsubsection{Power Spectral Density}\label{sec:psd1}

We further computed the PSD from the simulated HBABP dynamics at varied $V$ and compared them with that of the FDT prediction for passive Brownian diffusion, which is given by $ D_{\mathrm{T}} / \pi^2 f^2$. The PSDs match the FDT prediction only at higher frequencies and deviate from it where the activity dominates over spontaneous Brownian dynamics, as shown in Fig.~\ref{fig:MSDPSD1}(b). At lower frequencies, the PSDs exhibit saturation that increases with $V$, manifesting an activity-dependent effective confinement. The deviation of the PSDs from the FDT prediction and the variation in the plateau values with $V$ corroborate the activity-governed HBABP dynamics in this regime.

\subsection{Regime - II: $\tau_{\mathrm{R}} \gg \tau_k$}\label{sec:reg2}

This characteristic regime is commonly recognized as strongly active, i.e., the one that manifests the non-equilibrium dynamics of an HBABP, considering the bimodal (annular in 2D) position distribution in the steady state~\cite{pototskyActiveBrownianParticles2012, takatoriAcousticTrappingActive2016, basuLongtimePositionDistribution2019, malakarSteadyStateActive2020, chaudhuriActiveBrownianParticle2021, Santra2021, schmidtNonequilibriumPropertiesActive2021, buttinoniActiveColloidsHarmonic2022, Baldovin2023}. Following the same analyses as in the previous section, we show that the activity of an HBABP in this regime is balanced by the restoring force field at a radial distance $r$. Consequently, the resultant dynamics becomes devoid of residual activity and exhibits features similar to those of the passive counterpart, i.e., HBBP dynamics.

\begin{figure*}[ht]
	\centering
	\includegraphics[width=175mm]{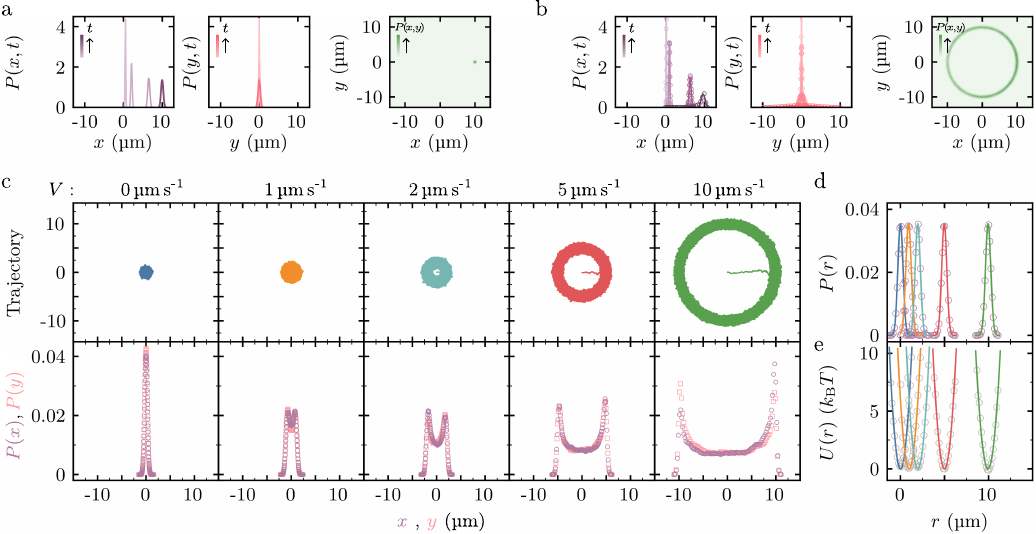}
	\caption{Position distributions in regime - II, where $\tau_{\mathrm{R}} \gg \tau_{k}$ ($\tau_{\mathrm{R}}/\tau_{k} = 100$; $\tau_{\mathrm{R}} =$ \SI{100}{\s}, $\tau_{k} =$ \SI{1}{\s}). (a)~Time evolution of $P(x, t)$ (left) and $P(y, t)$ (middle) from analytical predictions are shown by color-coded curves for $V = \SI{10}{\micro\meter\per\second}$. The 2D position distribution $P (x, y)$ prior to reaching the steady-state is presented as a density plot (right). (b)~Position distributions obtained from the numerically simulated dynamics of an HBABP are exhibited in the same fashion as those in (a). (c)~The simulated trajectories of \SI{E5}{\s} duration, generated at \SI{E3}{\Hz}, for five different propulsion speeds, from $V = 0$ (HBBP) to \SI{10}{\micro\meter\per\second}, are shown in blue, orange, cyan, red, and green (top). The bottom panel displays the corresponding 1D steady-state distributions $P(x)$ (circles) and $P(y)$ (squares). (d)~$P(r)$ (open circles) at these five $V$ values, computed from the simulated trajectories, are shown with Gaussian fitting (color-coded solid lines) following Eq.~\ref{eq:PrSS2} (e) Corresponding $U(r)$ (open circles), obtained by Boltzmann inversion of the respective $P(r)$, are exhibited with fitting to $(1/2)\,k_{\mathrm{eff}}\,(r - r_{\mathrm{c}})^2$ (color-coded solid lines), where $k_{\mathrm{eff}}$ remains unchanged and $r_{\mathrm{c}}$ increases linearly with $V$, in accordance with Eq.~\ref{eq:k_eff2}.}%
	\label{fig:PosDist2}
\end{figure*}

\subsubsection{Position distribution}\label{sec:pos2}

In this regime, the peak of the Gaussian position distribution along $\hat{x}$, the initial direction of propulsion, i.e., $P ( x; t )$, shifts progressively outward until it reaches $x = V \tau_{k}$ at $t \gg \tau_{k}$, as given by Eq.~\ref{eq:PxSS2} and shown in Fig.~\ref{fig:PosDist2}(a). The position distribution along the orthogonal direction $P (y; t)$ remains Gaussian, similar to that of an HBBP (Eq.~\ref{eq:PySS2}). At $t \gg \tau_{\mathrm{R}}$, the steady-state position distribution becomes isotropic, i.e., circularly symmetric, with the peak at a radially displaced position $ r = V \tau_{k}$, as expressed by Eq.~\ref{eq:PrSS2}. This annular-shaped position distribution appears bimodal and is composed of two symmetrically placed Gaussians when projected onto a single dimension. The radial distance of these Gaussians increases linearly with the propulsion speed $V$, whereas the variance remained unaltered, as demonstrated by the solid lines in Fig.~\ref{fig:PosDist2}(d). This indicates that the ABP is pushed to a radial distance by propulsion, where it is counterbalanced by a radially inward restoring force, and the resultant dynamics does not show any signature of activity. Thus, the HBABP dynamics effectively reduces to that of an HBBP in an apparently shifted harmonic well.

The time evolution of the numerically simulated $P (x; t)$ and $P (y; t)$, as exhibited in Fig.~\ref{fig:PosDist2}(b), reflects our analytical predictions. Simulated trajectories move away from the center and equilibrate at a radial distance, which eventually covers an annular region in the steady-state (Supp.~Video~\href{https://drive.google.com/file/d/16bpfX74Cn7hS5IDzzbZ9Xkn4gPCN1FPX/view?usp=drive_link}{2}). Typical annularly confined steady-state trajectories and the corresponding bimodal position distributions $P (x)$ and $P (y)$ at five different propulsion speeds are shown in Fig.~\ref{fig:PosDist2}(c). The increasing radius of the bound trajectories and the consequent outward shift of the peaks of the position distributions follow the analytical predictions. Furthermore, the radial distributions $P (r)$ at varied $V$ exhibit unaltered variance (Fig.~\ref{fig:PosDist2}(d)), in complete agreement with the corresponding analytical forms (Eq.~\ref{eq:PrSS2}), thereby corroborating the absence of any residual activity in the resultant HBABP dynamics.

Our experimental observations of optically trapped active Janus colloids also verify these results~\cite{Halder2026, Halder2025a}.

\subsubsection{Residual velocity}\label{sec:resv2}

\begin{figure}[ht]
	\centering
	\includegraphics[width=82mm]{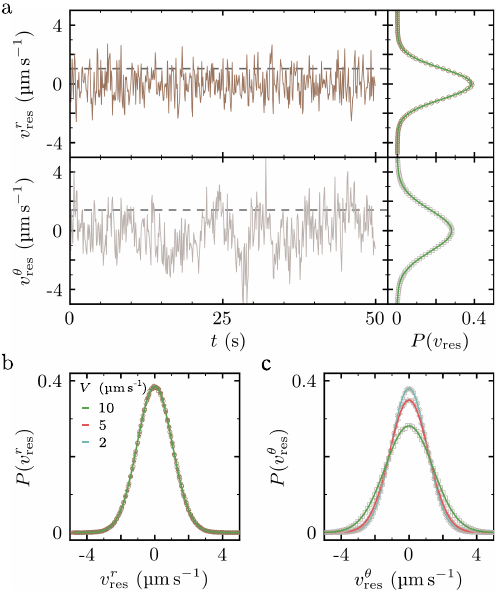}
	\caption{Residual velocity distributions of the HBABP for $\tau_{\mathrm{R}} \gg \tau_{k}$ ($\tau_{\mathrm{R}}/\tau_{k} = 100$; $\tau_{\mathrm{R}} =$ \SI{100}{\s}, $\tau_{k} =$ \SI{1}{\s}). (a) Short time series segments (\SI{50}{\s}) of $v_{\mathrm{res}}^{r}$ (top left) and $v_{\mathrm{res}}^{\theta}$ (bottom left), computed from the simulated trajectory at $V =$ \SI{10}{\micro\meter\per\second}, are shown, where the dashed lines represent the rms values. The corresponding probability distributions are exhibited on the right (circles and squares) along with the fitting to zero-mean Gaussian (solid lines). (b, c) Probability distributions $P(v_{\mathrm{res}}^{r})$ and $P(v_{\mathrm{res}}^{\theta})$ are shown for $V =$ 2, 5, and \SI{10}{\micro\meter\per\second}.}%
	\label{fig:ResVel2}%
\end{figure}

The residual velocity components $v_{\mathrm{res}}^r (t)$ and $v_{\mathrm{res}}^{\theta} (t)$ of the HBABP were obtained from the simulated trajectories at various propulsion speeds, $V$. For $\tau_{\mathrm{R}} \gg \tau_k$, the distributions of both $v_{\mathrm{res}}^r$ and $v_{\mathrm{res}}^{\theta}$ are Gaussian with significantly smaller rms values ($\sim$ \SI{1}{\micro\meter\per\second}) than the propulsion speed (\SI{10}{\micro\meter\per\second}), as exhibited in Fig. \ref{fig:ResVel2}(a). While the position distributions of $v_{\mathrm{res}}^r$ do not show any change with varying $V$, as the radial component of the active motion is steadily balanced by the restoring force field (Fig. \ref{fig:ResVel2}(b)), $P\left( v_{\mathrm{res}}^{\theta}\right)$ marginally widens with increasing propulsion speed (Fig. \ref{fig:ResVel2}(c)). Moreover, all the rms values of the residual velocities $v_{\mathrm{res}}$ at various $V$ are the same as the rms velocity of an HBBP, i.e., an HBABP with $V$ = 0~\cite{Halder2025a}. These results demonstrate that the resultant HBABP dynamics in this regime is devoid of activity, being similar to that of an HBBP, irrespective of the value of V, and are in stark contrast to the bimodal distributions with large rms values of the residual velocity components at $\tau_{\mathrm{R}} \ll \tau_k$.

Here, the restoring force completely balances $v_{\mathrm{res}}^r$, which is the principal component of the active propulsion, at a radial distance, and the ABP exhibits HBBP-like bound passive dynamics along the radial direction while remaining confined in an annular region. Our experimental results, where the distributions of both $v_{\mathrm{res}}^r (t)$ and $v_{\mathrm{res}}^{\theta} (t)$ are Gaussians with rms values that are substantially smaller than the propulsion speed and add up to only the Brownian velocity contribution, i.e., $v_{\mathrm{HBBP}}$ at $\tau_{\mathrm{R}} > \tau_k$, corroborate the same conclusion~\cite{Halder2025a}.

Therefore, the residual velocity analyses establish that, despite having a bimodal position distribution, which is commonly recognized as a far-from-equilibrium characteristic, the resultant HBABP dynamics in this regime are effectively reduced to those of an HBBP at a radially displaced position $r = V\tau_k$.

\subsubsection{Mean Square Displacement}\label{sec:msd2}

At $\tau_{\mathrm{R}} \gg \tau_k$, the analytical closed-form MSD is given by Eq.~\ref{eq:MSD2}. The MSD is the sum of two similar terms, where it grows linearly and eventually saturates at timescales $\tau_k$ and $\tau_{\mathrm{R}}$, respectively. At a shorter time, $\tau \gg \tau_k$, the ABP equilibrates in the harmonic well, which is expressed by the first term that is the same as the HBBP MSD, and the second term signifies the radial displacement of the equilibration position, i.e., $\Delta r = V \tau_k$. Thus, the MSD expresses HBBP-like dynamics, although at a displaced position. 

\begin{figure}[ht]
	\centering
	\includegraphics[width=68mm]{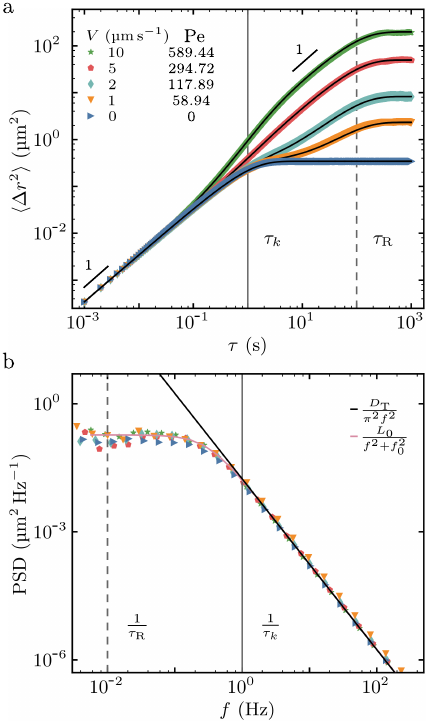}
	\caption{MSD and PSD of the HBABP in regime - II ($\tau_{\mathrm{R}}/\tau_{k} = 100$; $\tau_{\mathrm{R}} =$ \SI{100}{\s}, $\tau_{k} =$ \SI{1}{\s}). (a) MSDs, $\langle \Delta r^2 \rangle$, (solid symbols) for five different propulsion speeds, from $V = 0$ (HBBP) to \SI{10}{\micro\meter\per\second}, are shown. Their fitting with Eq.~\ref{eq:MSDGen} are presented as black solid lines. The respective $V$ and $\mathrm{Pe}$ values are listed in the inset. The characteristic timescales $\tau_{\mathrm{R}}$ and $\tau_{k}$ are indicated by dashed and solid vertical lines, respectively, with short black lines denoting slopes of 1. (b) Corresponding PSDs, calculated from the trajectory segments within thin radial strips along the positive $x$- and $y$-axes, are exhibited. Solid lines indicate the Lorentzian fit (Eq.~\ref{eq:PSD_HBBP}) and the FDT prediction for a free Brownian particle, given by $D_{\mathrm{T}} / \pi^2 f^2$. The vertical lines mark the characteristic frequencies $1/\tau_{k}$ (solid) and $1/\tau_{\mathrm{R}}$ (dashed).}%
	\label{fig:MSDPSD2}%
\end{figure}

The simulated MSDs at various $V$ values are shown in Fig.~\ref{fig:MSDPSD2}(a) along with their fit to Eq.~\ref{eq:MSDGen}. They exhibit two linear regimes at $\tau < \tau_k$ and at $\tau_k < \tau < \tau_{\mathrm{R}}$, which saturate at $\tau > \tau_k$ and $\tau > \tau_{\mathrm{R}}$, respectively, validating our analytical predictions. Both plateaus are apparent at lower $V$ values, whereas the first plateau is suppressed by the subsequent linear growth for stronger propulsion. These characteristic features in the MSD, signifying HBBP-like dynamics that maintains a radial distance at steady-state, are also observed in our experimental observations with an optically trapped active Janus colloid at $\tau_{\mathrm{R}} > \tau_k$~\cite{Halder2025a}.

\subsubsection{Effective harmonic confinement}\label{sec:keff2}

\begin{figure}[ht]
	\centering
	\includegraphics[width=65mm]{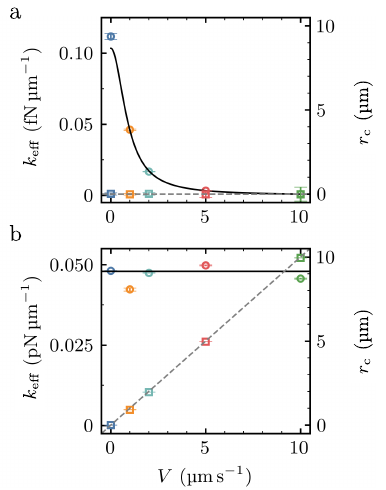}
	\caption{Dependence of the effective harmonic confinement on propulsion speed, $V$. The variations in the stiffness $k_{\mathrm{eff}}$ (left axis) and radial distance of the center $r_{\mathrm{c}}$ (right axis) with $V$ are shown for (a) regime - I and (b) regime - II. Color-coded open circles ($k_{\mathrm{eff}}$) and squares ($r_{\mathrm{c}}$) with error bars represent the values obtained from the fitting of the effective confining potentials (Fig. \ref{fig:PosDist1}(e) and \ref{fig:PosDist2}(e)), and the solid and dashed lines indicate theoretical predictions for $k_{\mathrm{eff}}$ and $r_{\mathrm{c}}$, given by (a) Eq. \ref{eq:k_eff1}, and (b) Eq. \ref{eq:k_eff2}.}%
	\label{fig:k_eff}%
\end{figure}

In this regime, the stiffness of the effective harmonic confinement $k_{\mathrm{eff}}$ and its position $r_{\mathrm{c}}$, representing the steady-state bound dynamics of the HBABP, are expressed by Eq.~\ref{eq:k_eff2}. Here, the activity of the ABP does not alter the effective stiffness of the harmonic confinement, verifying that the resultant HBABP dynamics is essentially reduced to HBBP motion, where the activity is used to displace the center of the confinement to a radial distance $r_{\mathrm{c}} = V \tau_k$. The effective harmonic potentials $U (r)$ are obtained from the simulated steady-state position distribution $P (r)$ by Boltzmann inversion at various $V$ values, as shown in Fig.~\ref{fig:PosDist2}(e). The form of $U (r)$ remains unchanged, whereas it moves progressively radially outward with increasing $V$, validating our theoretical description of the effective confinement. Additionally, the variation in $r_{\mathrm{c}}$ and $k_{\mathrm{eff}}$ with $V$, obtained from the fitting of the corresponding $U (r)$, match excellently with the analytical prediction (Eq.~\ref{eq:k_eff2}), as shown in Fig. \ref{fig:k_eff}. The same variation in the effective confinement, as described by Eq.~\ref{eq:k_eff2}, has also been experimentally verified by optically trapping an active Janus colloid~\cite{Halder2026}.

\subsubsection{Power Spectral Density}\label{sec:psd2}

The PSDs computed from the simulated radial HBABP dynamics (considering segments of a trajectory from within a radial trip) in this regime at various $V$ values are compared with those of the FDT prediction for passive Brownian diffusion ($D_{\mathrm{T}} / \pi^2 f^2$) and for an HBBP, given by a Lorentzian $ L_0 / (f^2+f_0^2)$ (Eq.~\ref{eq:PSD_HBBP}), in Fig.~\ref{fig:MSDPSD2}(b). All PSDs, irrespective of $V$, collapse on the PSD of the corresponding HBBP, i.e., with $V$ = 0, and match the FDT prediction for free Brownian diffusion at frequencies higher than the corner frequency $f_0 = (2 \pi \tau_k)^{-1}$. These results corroborate that the radial HBABP dynamics in this regime is fully devoid of activity, essentially being reduced to that of the corresponding HBBP. Consequently, the resultant HBABP motion, which is predominantly along the radial direction with a slow azimuthal drift at long times ($t < \tau_{\mathrm{R}}$), does not show any significant signature of the activity. Here, the activity is counteracted and compensated for by harmonic confinement at a radial distance, where the HBABP exhibits HBBP-like dynamics.

\subsection{Intermediate regime: $\tau_{\mathrm{R}} = \tau_k$}\label{sec:reg3}

The intermediate regime between the two,$\tau_{\mathrm{R}} \ll \tau_k$ and $\tau_{\mathrm{R}} \gg \tau_k$, which are characterized by distinctly different dynamical features, remains to be analyzed. However, most of the analytical descriptions developed for these limiting cases break down at $\tau_{\mathrm{R}} = \tau_k$, i.e., at the crossover. Here, we examine the dynamical properties of an HBABP in this intermediate regime.

\begin{figure*}[ht]
	\centering
	\includegraphics[width=136mm]{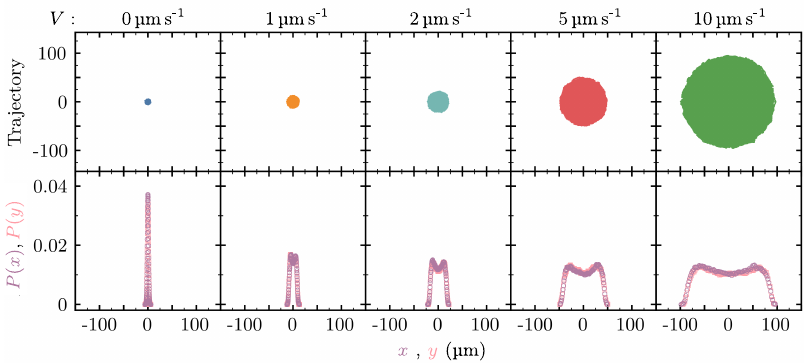}
	\caption{Simulated trajectories and position distributions of an HBABP in the intermediate regime with $\tau_k = \tau_{\mathrm{R}} = \SI{10}{s}$. Numerically simulated trajectories of \SI{E5}{\s} duration are shown for varied propulsion speeds $V$, starting from 0 (HBBP) to \SI{10}{\micro\meter\per\second} in blue, orange, cyan, red, and green (top panel). The corresponding 1D steady-state position distributions $P(x)$ (circles) and $P(y)$ (squares) are displayed in the bottom panel. }%
	\label{fig:PosDist3}%
\end{figure*}

\subsubsection{Position distribution}\label{sec:pos3}

In this regime, the positional distribution does not have an analytical closed-form solution. We simulated the HBABP dynamics with $\tau_{\mathrm{R}} = \tau_k$ = \SI{10}{\second}. The simulated trajectories are bound, space-filling, and center-avoiding (\href{https://drive.google.com/file/d/1EToFYCh39pHFejzeOe8I-CffrSpCgkeZ/view?usp=drive_link}{Supp. Video 3}), as shown in Fig.~\ref{fig:PosDist3}. The spread of the bound trajectories and, consequently, the widths of the weakly bimodal position distributions $P (x)$ and $P (y)$ increase with the propulsion speed $V$, whereas the dip at the middle of $P (x)$ and $P (y)$ indicates that the trajectories are center-avoiding. These features are intermediate to, but distinctly different, from the Boltzmann-like position distribution at $\tau_{\mathrm{R}} \ll \tau_k$ and annularly confined trajectories with strongly bimodal 1D position distributions at $\tau_{\mathrm{R}} \gg \tau_k$.

This intermediate behavior can be explained in terms of the competing timescales $\tau_{\mathrm{R}}$ and $\tau_k$. When these timescales are equal, the propulsion direction persists long enough for the restoring force field to partially balance the radial component of the active velocity at a radial distance from the center. However, the reorientation of propulsion is not sufficiently slow for the ABP to equilibrate at the displaced radial position and perform a slow azimuthal motion at long time to eventually exhibit an annularly confined trajectory in the steady-state. Therefore, the HBABP explores a broad region around the center, resulting in a weakly bimodal position distribution that widens with $V$, signifying the presence of moderately diminished propulsion, i.e., weak residual activity.

\subsubsection{Residual velocity}\label{sec:resv3}

\begin{figure}[ht]
	\centering
	\includegraphics[width=85mm]{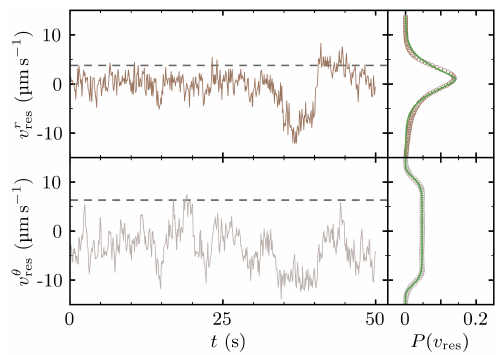}
	\caption{Residual velocity distributions for the intermediate regime ($\tau_k = \tau_{\mathrm{R}} = \SI{10}{s}$). A short segment (\SI{50}{s}) of the time series of the radial and azimuthal components of the residual velocity, $v_{\mathrm{res}}^r$ and $v_{\mathrm{res}}^{\theta}$, respectively, computed from the simulated HBABP trajectory with $V$ = \SI{10}{\micro\meter\per\second}, are shown along with the corresponding probability distributions on the right. The respective rms values are marked by dashed horizontal lines in the time-series plots. $P(v_{\mathrm{res}}^r)$ (circles) is fitted to a super-Gaussian (Eq.~\ref{eq:SuperGauss}) with $n$ = 1.44 (solid line) and $P(v_{\mathrm{res}}^{\theta})$ (squares) with $n$ = 9.3, which represents a flat-topped Gaussian (solid line).}%
	\label{fig:ResVel3}%
\end{figure}

We further computed the residual velocity components $v_{\mathrm{res}}^r (t)$ and $v_{\mathrm{res}}^{\theta} (t)$ of the HBABP from the simulated trajectories with $\tau_{\mathrm{R}} = \tau_k$. Here, the probability distributions of both $v_{\mathrm{res}}^r$ and $v_{\mathrm{res}}^{\theta}$ are neither bimodal nor purely Gaussian, rather described by super-Gaussian functions, given by,
\begin{equation}
	f(x) = A \exp\left( - \left( |x - \mu| / w \right)^n \right),
	\label{eq:SuperGauss}
\end{equation}
\noindent where $w$ is the width and $n$ is the shape parameter. While $P(v_{\mathrm{res}}^r)$ fits well to a peaked profile that is close to a Gaussian with $n$ = 1.44, $P(v_{\mathrm{res}}^{\theta})$ fits to a flat-topped profile with $n$ = 9.3, as shown in Fig.~\ref{fig:ResVel3}. The shapes of the probability distributions indicate that the restoring force cannot fully balance the radial component of the active velocity $V^r$ but diminishes it substantially, whereas the azimuthal component of propulsion $V^{\theta}$ remains unimpeded, and $v_{\mathrm{res}}^{\theta}$ dominates the resultant dynamics, being the principal component of the residual velocity. This is further corroborated by the rms values of $v_{\mathrm{res}}$ components (Fig.~\ref{fig:ResVel3}).

\subsubsection{Mean Square Displacement}\label{sec:msd3}

The generic MSD expression, which is given by Eq.~\ref{eq:MSDGen}, diverges at $\tau_{\mathrm{R}} = \tau_k$. We derived the closed-form analytical solution for MSD considering $\tau_{\mathrm{R}} = \tau_k = \tau_{\mathrm{eq}}$ (Appendix~\ref{appendix:c}), and it is expressed as,
\begin{equation}
	\begin{split}
		\langle \Delta r^2(\tau)\rangle = & \frac{4k_{\mathrm{B}} T}{k}  \left(1- e^{- \tau/ \tau_{\mathrm{eq}}} \right) \\
		& \quad + V^2 \tau^2_{\mathrm{eq}} \left[ 1- e^{- \tau/ \tau_{\mathrm{eq}}}\left(1 - \frac{\tau}{\tau_{\mathrm{eq}}} \right) \right].
	\end{split}
	\label{eq:MSDEqual}
\end{equation}
The second term represents ballistic growth in the MSD at the intermediate time-lags, accounting for the residual activity until it saturates at $V^2 \tau_{\mathrm{eq}}^2$, while the first term is the contribution from the corresponding HBBP dynamics.

\begin{figure}[ht]
	\centering
	\includegraphics[width=72mm]{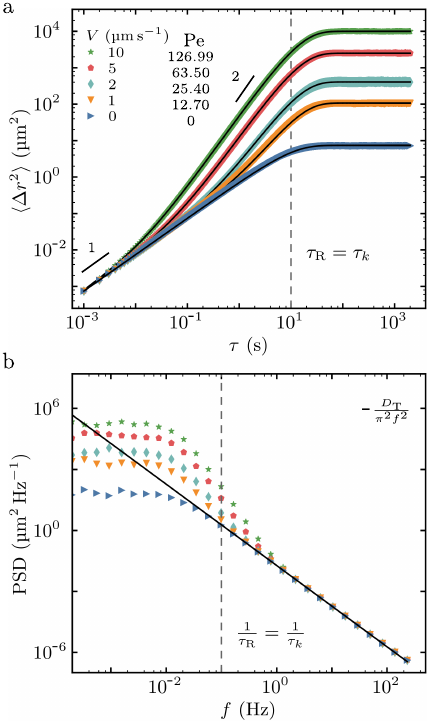}
	\caption{MSD and PSD of HBABP dynamics, where $\tau_k = \tau_{\mathrm{R}} = \SI{10}{s}$. (a) MSDs computed from simulated HBABP dynamics for five different propulsion speeds, from $V$ = 0 (HBBP) to \SI{10}{\micro\meter\per\second}, are shown with color-coded symbols. The respective $V$ and $\mathrm{Pe}$ values are given in the inset. Solid black lines represent fits with the analytical solution (Eq.~\ref{eq:MSDEqual}). The vertical dashed line marks both $\tau_{\mathrm{R}}$ and $\tau_k$, which are equal, and the apparent slopes of 1 and 2 are indicated by short black straight lines as visual guides. (b) Corresponding PSDs are compared with the FDT prediction for a free Brownian particle (black straight lines); the vertical line marks the characteristic frequency, $1/\tau_{\mathrm{R}} \equiv 1/\tau_k$.}%
	\label{fig:MSDPSD3}%
\end{figure}

All the simulated MSDs with $\tau_{\mathrm{R}} = \tau_k$ = \SI{10}{\second} at varied $V$ fit excellently with Eq.~\ref{eq:MSDEqual}, as shown in Fig.~\ref{fig:MSDPSD3}(a). Here, the MSDs grow uniformly as $\tau^2$ at the intermediate time-lags before reaching a plateau, unlike the cases with well-separated $\tau_{\mathrm{R}}$ and $\tau_k$ values. The MSD in this regime is transitional between the two above-mentioned cases, i.e., activity-dominated dynamics at $\tau_{\mathrm{R}} \ll \tau_k$ and HBBP-like effective motion at $\tau_{\mathrm{R}} \gg \tau_k$.

\subsubsection{Power Spectral Density}\label{sec:psd3}

We further computed PSDs from the simulated HBABP dynamics at $\tau_{\mathrm{R}} = \tau_k$ = \SI{10}{\second} and compared them with the FDT prediction for passive Brownian diffusion ($ D_{\mathrm{T}} / \pi^2 f^2 $), as exhibited in Fig.~\ref{fig:MSDPSD3}(b). Similar to the activity-governed regime at $\tau_{\mathrm{R}} \ll \tau_k$, the PSDs deviate from the FDT prediction at progressively larger frequencies with increasing $V$, consistent with the presence of moderate residual activity at $\tau_{\mathrm{R}} = \tau_k$.

\section{Conclusions}\label{sec:conclusions}

This study provides a comprehensive analysis of HBABP dynamics with analytical calculations and numerical simulations in two characteristically disparate regimes, observed at $\tau_{\mathrm{R}} \ll \tau_k$ and $\tau_{\mathrm{R}} \gg \tau_k$, irrespective of the propulsion speed~\cite{Halder2025a}, and thus identifies the signatures of activity in the resultant motion. The steady-state position distributions are Boltzmann-like at $\tau_{\mathrm{R}} \ll \tau_k$ and bimodal at $\tau_{\mathrm{R}} \gg \tau_k$, reflecting equilibrium-like and strongly out-of-equilibrium characteristics, respectively. Accordingly, the corresponding regimes are commonly referred to as the passive and active regimes~\cite{pototskyActiveBrownianParticles2012, takatoriAcousticTrappingActive2016, basuLongtimePositionDistribution2019, malakarSteadyStateActive2020, chaudhuriActiveBrownianParticle2021, Santra2021, schmidtNonequilibriumPropertiesActive2021, buttinoniActiveColloidsHarmonic2022, Baldovin2023}.

However, the variance of the Boltzmann-like steady-state position distribution at $\tau_{\mathrm{R}} \ll \tau_k$ increases monotonically with the propulsion speed, signifying an activity-dominated resultant bound dynamics in this regime. The resultant or residual velocity, $v_{\mathrm{res}}$, analysis makes this more apparent. Both the radial and azimuthal components of $v_{\mathrm{res}}$ exhibit bimodal distributions, with the peaks symmetrically positioned at nearly the corresponding rms values, which are significantly large and add up in quadrature to be approximately equal to the propulsion speed. This indicates that the HBABP retains all of its active motion, which governs the resultant dynamics even under harmonic confinement. The MSDs in this regime show a clear ballistic rise at intermediate times, and the plateau values increase almost quadratically with the propulsion speed, reflecting activity-governed resultant motion. Furthermore, the effective harmonic confinement describing the HBABP at steady state becomes weaker with enhanced activity, and the PSDs of the resultant dynamics deviate strongly from the FDT prediction for passive Brownian dynamics, corroborating this conclusion.

In the other regime, at $\tau_{\mathrm{R}} \gg \tau_k$, the peaks of the bimodal position distribution appear at a distance where the propulsion is balanced by the restoring force, and the width of the peaks remains the same as that of the corresponding HBBP, regardless of the propulsion speed. This indicates that the resultant HBABP dynamics essentially reduces to an HBBP-like motion at a displaced position. Both components of $v_{\mathrm{res}}$ exhibit Gaussian probability distributions with significantly smaller rms values, similar to those of the HBBP \cite{Halder2025a}, verifying our conclusion. Here, the MSDs are composed of two distinct contributions: one from the corresponding HBBP dynamics and the other from its displacement from the center. Moreover, the MSDs in this regime are significantly ($\sim$ 100 times) smaller than those with the same propulsion speed in regime - I. The strength of the effective harmonic confinement $k_{\mathrm{eff}}$ remains the same as that of the corresponding HBBP, being independent of the propulsion speed $V$, whereas the radial distance of its center $r_{\mathrm{c}}$ increases linearly with $V$. Finally, the PSDs for all values of $V$ collapse into a single Lorentzian that represents the PSD of the corresponding HBBP. These findings further validate that the resultant HBABP dynamics in this regime is devoid of activity, which is consumed in displacing the confined dynamics away from the center.

For completeness, we also analyze the HBABP dynamics in the crossover regime, where $\tau_{\mathrm{R}} = \tau_k$. Here, the dynamical properties are intermediate to those in the two characteristic regimes discussed above. These analyses demonstrate that although the steady-state position distributions apparently contradict the probability distributions of the residual velocity components in those two regimes, both distributions signify the same crossover, which is from activity-dominated dynamics at $\tau_{\mathrm{R}} \ll \tau_k$ to activity-depleted HBBP dynamics at $\tau_{\mathrm{R}} \gg \tau_k$, where the activity is neutralized by the restoring force at a radially displaced position. At $\tau_{\mathrm{R}} \ll \tau_k$, the propulsion direction of the ABP changes so rapidly that it never settles down at a radial position where the radial component of the propulsion is compensated by the restoring force field. Consequently, the ABP retains its active motion and remains bound by the harmonic well, as the propulsion can only take it to a certain radial distance, working against the restoring force. Therefore, a harmonically bound ABP is less likely to be found at a longer radial position. This leads to a Boltzmann-like position distribution without a long deviation, which is observed in the case of free ABP beyond the persistence time \cite{Bechinger2016, Fodor2018, basuActiveBrownianMotion2018, basuLongtimePositionDistribution2019}.

We obtained closed-form position distributions relying on the fact that the composite noise is delta-correlated; this condition is satisfied only in the two limiting regimes. For an arbitrary $\tau_{\mathrm{R}}/\tau_k$ ratio, the steady-state distribution remains analytically intractable, although exact moments and moment-expansion solutions have been obtained~\cite{chaudhuriActiveBrownianParticle2021, caraglioAnalyticSolutionActive2022}. However, other analyses, such as residual velocity decomposition, effective confinement description, and variations in MSD and PSD, which are used in this study, remain applicable for any $\tau_{\mathrm{R}}/\tau_k$ ratio. The effective harmonic confinement parameters, $k_{\mathrm{eff}}$ and $r_{\mathrm{c}}$, obtained consistently from both the steady-state position distributions and MSDs provide a complete steady-state description of an HBABP, unlike the effective temperature framework~\cite{szamelSelfpropelledParticleExternal2014}. Our study further establishes that $\mathrm{Pe}$ cannot be a reliable measure of residual activity in the case of confined ABPs.

The experimental results validating this dynamical crossover from $\tau_{\mathrm{R}} \ll \tau_k$ to $\tau_{\mathrm{R}} \gg \tau_k$ are presented in the companion Letter \cite{Halder2025a}. Analyzing the signature of activity in the resultant motion of other active systems governed by the Ornstein–Uhlenbeck process or run-and-tumble dynamics under harmonic or anharmonic confinement~\cite{solonActiveBrownianParticles2015, dasConfinedActiveBrownian2018, GarciaMillan2021, Semeraro2023, Dutta2024, Sun2025}, or in complex environments, such as position-dependent confinement with spatially varying restoring timescales, and viscoelastic media that introduce additional timescales because of memory~\cite{Khan2014a, Sprenger2022}, remains open directions. These analyses can also be applied to identify the signature of residual activity in the collective systems of active particles under crowding, such as in active glasses \cite{Ni2013, Berthier2013, Janssen2019}.

\vspace{1em}
\section*{Acknowledgements}
The authors acknowledge the Science and Engineering Research Board (SERB), Govt.\ of India, for supporting this work through a Core Research Grant (CRG/2020/002723), and the PARAM Sanganak computing facility at the Computer Center, IIT Kanpur, for the numerical simulations. MK thanks Abhik Basu, Ambarish Ghosh, and Sriram Ramaswamy for fruitful discussions and critical reading of the manuscript.

\section*{Disclosures}
The authors declare no competing interest.

\section*{Data availability}
All data required to reach the conclusions of this study are presented in the manuscript.

\section*{Author contributions}
All authors contributed to the conception and design of the research. S.H. performed the analytical calculations and numerical simulations. S.H. and M.K. interpreted data and wrote the manuscript. M.K. supervised the project.

\clearpage

\appendix
\makeatletter
\@addtoreset{figure}{section}
\@addtoreset{table}{section}
\makeatother
\renewcommand{\thefigure}{\thesection\arabic{figure}}
\renewcommand{\thetable}{\thesection\arabic{table}}
\vspace{3em}
\begin{center}
\textbf{\large \noindent Appendix} %
\end{center}

\section{Mean values and correlation functions}\label{appendix:a}

The probability of $\phi(t)$ is given by~\cite{tenhagenBrownianMotionSelfpropelled2011},
\begin{equation}
	P\left( \phi, \phi_0 ; t \right) = \frac{1}{\sqrt{4\pi D_{\mathrm{R}} t}} \exp \left( -\frac{(\phi - \phi_0)^2}{4D_{\text{R}} t}\right).
	\label{eq:PhiPDF}
\end{equation}%
Since the Brownian velocities have zero mean, the ensemble averages of $A^x (t)$ and $A^y (t)$ reduce to the active contributions alone:
\begin{align}
   \left\langle A^x (t) \right\rangle & = \left\langle v^x_{\mathrm{B}} (t)\right\rangle + \left\langle V \cos(\phi(t))\right\rangle = V \left\langle \cos(\phi(t))\right\rangle, \; \text{and} \label{eq:AvgAx} \\
   \left\langle A^y (t) \right\rangle & = \left\langle v^y_{\mathrm{B}} (t)\right\rangle + \left\langle V \sin(\phi(t))\right\rangle = V \left\langle \sin(\phi(t))\right\rangle. \label{eq:AvgAy}
\end{align}%
Using Eq.~\ref{eq:PhiPDF}, $\left\langle \cos (\phi(t)) \right\rangle = e^{-D_{\mathrm{R}} t} \cos \phi_0$ and $\left\langle \sin (\phi(t)) \right\rangle = e^{-D_{\mathrm{R}} t} \sin \phi_0$. Substituting these into Eqs.~\ref{eq:AvgAx} and~\ref{eq:AvgAy} gives
\begin{align}
    \left\langle A^x (t) \right\rangle & = V \cos(\phi_0)\, e^{-t/\tau_{\mathrm{R}}},\; \text{and}\\
    \left\langle A^y (t) \right\rangle & = V \sin(\phi_0)\, e^{-t/\tau_{\mathrm{R}}}.
\label{eq:Amean}
\end{align}%
In the autocorrelation of $A^x (t)$, only the autocorrelations of the Brownian velocity $v^x_{\mathrm{B}} (t)$ and active velocity component $V \cos (\phi (t))$ survive because $v^x_{\mathrm{B}} (t)$ is uncorrelated to the orientation of the particle $\phi (t)$, and hence, to $\cos (\phi (t))$. Therefore,
\begin{align}
 \langle A^x(t_1)A^x(t_2)\rangle & = 2 D_{\mathrm{T}} \delta(t_1 - t_2) \nonumber \\
								 & \qquad + V^2 \langle \cos(\phi(t_1)) \cos(\phi(t_2))\rangle, \label{eq: ACAx} \\
 \langle A^y(t_1)A^y(t_2)\rangle & = 2 D_{\mathrm{T}} \delta(t_1 - t_2) \nonumber \\
								 & \qquad + V^2 \langle \sin(\phi(t_1)) \sin(\phi(t_2))\rangle. \label{eq: ACAy}								 
\end{align}%
The autocorrelations of $\cos (\phi(t))$ and $\sin(\phi (t))$ can be evaluated for the given $P\left( \phi, \phi_0 ; t \right)$ (Eq.~\ref{eq:PhiPDF}), to obtain
\begin{align}
	& \left\langle \cos (\phi(t_1)) \cos (\phi(t_2)) \right\rangle \nonumber \\
	& = \frac{1}{2} e^{- \left| t_1 - t_2 \right| / \tau_{\mathrm{R}}} \left(1+ \cos(2 \phi_0) e^{- 4 \min(t_1, t_2) / \tau_{\mathrm{R}}} \right) , \; \text{and} \label{eq:ACcosphi} \\
	& \left\langle \sin (\phi(t_1)) \sin (\phi(t_2)) \right\rangle \nonumber \\
	& = \frac{1}{2} e^{- \left| t_1 - t_2 \right| / \tau_{\mathrm{R}}} \left(1- \cos(2 \phi_0) e^{- 4 \min(t_1, t_2) / \tau_{\mathrm{R}}} \right). \label{eq:ACsinphi}
\end{align}

The initial position $x(0) =x_0$ is obtained from Eq.~\ref{eq:x} at $t = 0$ as~\cite{doiSoftMatterPhysics2013}
\begin{align}
	 x_0 & = \int_{- \infty}^{0} e^{t^{\prime}/\tau_k} \left[ v^x_{\mathrm{B}}(t^{\prime}) +V \cos(\phi(t^{\prime}))\right] dt^{\prime} \notag \\
	     & = \int_{- \infty}^{0} e^{t^{\prime}/\tau_k} A^x(t^{\prime})\, dt^{\prime}. \label{eq:x0}
\end{align}%
The position autocorrelation function for the general case, where $\tau_{\mathrm{R}} \neq \tau_k$, is derived as
\begin{equation}
	\begin{split}
		& \langle x(t) x(0) \rangle \\
		& = \left\langle e^{-t / \tau_k} \int_{- \infty}^{t} e^{t_1/\tau_k} \left[ v^x_{\mathrm{B}}(t_1) +V\cos(\phi(t_1))\right] dt_1 \times \right.\\ 
		& \qquad \left. \int_{- \infty}^{0} e^{t_2/\tau_k} \left[ v^x_{\mathrm{B}}(t_2) +V\cos(\phi(t_2))\right] dt_2 \right\rangle \\
		& = \int_{-\infty}^{t} dt_1 \int_{-\infty}^{0} dt_2 \, e^{-(t - t_1 - t_2)/\tau_k} \langle v^x_{\mathrm{B}}(t_1) v^x_{\mathrm{B}}(t_2) \rangle \\
		& \quad + V^2 \int_{-\infty}^{t} dt_1 \int_{-\infty}^{0} dt_2 \, e^{-(t - t_1 - t_2)/\tau_k} \times \\
		& \hspace{3cm}  \left\langle \cos(\phi(t_1))\cos(\phi(t_2))\right\rangle \\                                                                                                                 
		& = \int_{-\infty}^{t} dt_1 \int_{-\infty}^{0} dt_2 \, e^{-(t - t_1 - t_2)/\tau_k} \langle v^x_{\mathrm{B}}(t_1) v^x_{\mathrm{B}}(t_2) \rangle \\ 
		& \quad + V^2 \biggl[  \int_{-\infty}^{t_1} dt_1 \int_{-\infty}^{0} dt_2 \, e^{-(t - t_1 - t_2)/\tau_k} \times  \\ 
		& \hspace{3cm} \left\langle \cos(\phi(t_1))\cos(\phi(t_2))\right\rangle_{t_2<t_1}  \\
		& \qquad  + \int_{t_1}^{t} dt_1 \int_{-\infty}^{0} dt_2 \, e^{-(t - t_1 - t_2)/\tau_k} \times \\
		&  \hspace{3cm}  \langle \cos(\phi(t_1))\cos(\phi(t_2))\rangle_{t_2 > t_1} \biggr]  \\                                                          
		& = \int_{-\infty}^{t} dt_1 \int_{-\infty}^{0} dt_2 \, e^{-(t - t_1 - t_2)/\tau_k} \frac{2k_{\mathrm{B}}T}{\gamma_T} \delta(t_1- t_2) \\
		& \quad + \frac{V^2}{2}\left[  \int_{-\infty}^{t_1} dt_1 \int_{-\infty}^{0} dt_2 \, e^{-(t - t_1 - t_2)/\tau_k} \, e^{-(t_1 - t_2)/ \tau_{\mathrm{R}}} \right. \\
		& \qquad \left. + \int_{t_1}^{t} dt_1 \int_{-\infty}^{0} dt_2 \, e^{-(t - t_1 - t_2)/\tau_k} \, e^{-(t_2 - t_1)/ \tau_{\mathrm{R}}} \right] \\ 
		& \quad + I_{1}  \cos2\phi_0 \\
		& = \frac{k_{\mathrm{B}} T}{k} e^{- t/ \tau_k} + \frac{V^2}{4} \left[  \frac{\tau^2_k \tau_{\mathrm{R}}}{\tau_{\mathrm{R}} + \tau_k} e^{- t/ \tau_k} - \frac{\tau^2_k\tau_{\mathrm{R}}}{\tau_{\mathrm{R}} - \tau_k} e^{- t/ \tau_k} \right.\\
		& \qquad \left. + \frac{ 2 \tau^2_k \tau^2_{\mathrm{R}}}{(\tau_{\mathrm{R}} - \tau_k)(\tau_{\mathrm{R}} + \tau_k)} e^{- t/ \tau_{\mathrm{R}}}\right]  \\ 
		& \quad +  I_{1}  \cos2\phi_0 \\
		& = \frac{k_{\mathrm{B}} T}{k} e^{- t/ \tau_k} + \frac{V^2}{2} \frac{\tau_{\mathrm{R}} \tau^2_k \left( \tau_{\mathrm{R}} e^{- t/ \tau_{\mathrm{R}}} - \tau_k e^{- t/ \tau_k}\right) }{ \left( \tau_{\mathrm{R}} + \tau_k\right) \left( \tau_{\mathrm{R}} - \tau_k\right) }  \\ 
		& \quad +  I_{1}  \cos2\phi_0
		\label{eq:MSDx1}
	\end{split}
\end{equation}%
where 
\begin{align}
I_1 & =  \frac{V^2}{2} \left[  \int_{-\infty}^{t_1} dt_1 \int_{-\infty}^{0} dt_2 \, e^{-(t - t_1 - t_2)/\tau_k} \, e^{-(t_1 + 3t_2)/ \tau_{\mathrm{R}}} \right. \notag \\
    & \left. + \int_{t_1}^{t} dt_1 \int_{-\infty}^{0} dt_2 \, e^{-(t - t_1 - t_2)/\tau_k} \, e^{-(3t_1 + t_2)/ \tau_{\mathrm{R}}} \right].
\end{align}%
By following similar steps, we obtain
\begin{equation}
	\langle x^2(t) \rangle  = \langle x^2(0) \rangle = \frac{k_{\mathrm{B}} T}{k} + \frac{V^2}{2} \frac{\tau_{\mathrm{R}} \tau^2_k }{ \left( \tau_{\mathrm{R}} + \tau_k\right) }  +  I_{2}  \cos2\phi_0 
	\label{eq:MSDx2}
\end{equation}
where 
\begin{align}
	I_2 & =  \frac{V^2}{2} \left[  \int_{-\infty}^{t_1} dt_1 \int_{-\infty}^{t} dt_2 \, e^{-(t - t_1 - t_2)/\tau_k} \, e^{-(t_1 + 3t_2)/ \tau_{\mathrm{R}}} \right. \nonumber \\
	    & \quad \left. + \int_{t_1}^{t} dt_1 \int_{-\infty}^{t} dt_2 \, e^{-(t - t_1 - t_2)/\tau_k} \, e^{-(3t_1 + t_2)/ \tau_{\mathrm{R}}} \right].
\end{align}%

\section{Position distribution}\label{appendix:b}
We used the following lemma~\cite{Chandrasekhar1943} to derive the position distributions $P(x, x_0; t)$ and $P(y, y_0; t)$ for the two extreme cases: $\tau_{\mathrm{R}} \ll \tau_k$ and $\tau_{\mathrm{R}} \gg \tau_k$.

\noindent\textit{Lemma:} If $R (t) = \int_{0}^{t}\psi(t^{\prime}) A_{\mathrm{B}}(t^\prime) dt^{\prime}$, where $A_{\mathrm{B}}(t)$ is a Gaussian random variable with mean $\langle A_{\mathrm{B}}(t) \rangle = 0$ and autocorrelation $\langle A_{\mathrm{B}}(t_1)A_{\mathrm{B}}(t_2) \rangle = 2q \delta(t_1 - t_2)$, then the probability distribution of $R (t)$ is given by,
\begin{equation}
	P(R; t) = \frac{1}{\sqrt{4\pi q\int_{0}^{t}\psi^2(t^{\prime})\, dt^{\prime}}}\exp \left( - \frac{|R|^2}{4q\int_{0}^{t}\psi^2(t^{\prime})\, dt^{\prime}}\right).
\end{equation}
\\
\noindent
\textbf{Case-1: $\boldsymbol{\tau_{\mathrm{R}} \ll \tau_k}$} \\
\\
\noindent This condition represents cases in which the harmonic confinement is very weak or the ABP has extremely fast orientational diffusion. At this limit, the directional correlation of the particle decays almost instantaneously compared to its equilibration in the potential well. Therefore, for all practical purposes, $\tau_{\mathrm{R}}$ can be considered insignificantly short, i.e., $\tau_{\mathrm{R}} \to$ 0. Consequently, the mean and autocorrelation of $A^x (t)$ and $A^y (t)$, given by Eq.~\ref{eq:Amean}  to Eq.~\ref{eq:ACsinphi}, are reduced to 
\begin{align}
	\left\langle A^x (t) \right\rangle & = \left\langle A^y (t) \right\rangle = 0, \quad \text{and} \nonumber \\ 
	\left\langle A^x(t_1)A^x(t_2) \right\rangle & = \left\langle A^y(t_1)A^y(t_2) \right\rangle \nonumber \\
											    & = \left[ 2 D_{\mathrm{T}} + V^2 \tau_{\mathrm{R}} \right]  \delta(t_1 - t_2)
	\label{AxyProp1}
\end{align}
This allows us to use the lemma in Eq.~\ref{eq:x}, where $\left(x - x_0 e^{-t/\tau_k} \right)$, $e^{(t^{\prime} - t) / \tau_k}$, and $A^x (t)$ correspond to $R (t)$, $\psi(t^{\prime})$, and $A_{\mathrm{B}}(t)$, respectively. Thus, using the following equivalences 
\begin{align}
	q & \equiv D_{\mathrm{T}} + \frac{1}{2} V^2 \tau_{\mathrm{R}}, \quad \text{and} \\
	\int_{0}^{t}\psi^2 (t^{\prime})\, dt^{\prime} & \equiv  \int_{0}^{t} e^{(t^{\prime} - t) / \tau_k}\, dt^{\prime} = \frac{\tau_k}{2} \left( 1 - e^{- 2t / \tau_k} \right),
\end{align}%
we obtain
\begin{equation}
	\begin{split}
		& P (x, x_0; t) \equiv P\left(x - x_0 e^{-t/\tau_k} \right) \\
		& = \frac{1}{\sqrt{2 \pi \left(D_{\mathrm{T}} + V^2 \tau_{\mathrm{R}}/2\right) \tau_k \left( 1 - e^{- 2t / \tau_k} \right)}} \times \\
		& \qquad \exp \left[ - \frac{\left(x - x_0 e^{-t/\tau_k} \right)^2}{2 \left(D_{\mathrm{T}} + V^2 \tau_{\mathrm{R}}/2\right) \tau_k \left( 1 - e^{- 2t / \tau_k} \right)}\right]. 
	\label{eq:Px}
	\end{split}
\end{equation}%
Denoting $\left(D_{\mathrm{T}} + V^2 \tau_{\mathrm{R}}/2\right)\tau_k \left( 1 - e^{- 2t / \tau_k} \right) = \frac{k_{\mathrm{B}}T}{k}\times$ \\ $\left(1 + \frac{\mathrm{Pe}^2}{2}\right) \left( 1 - e^{- 2t / \tau_k} \right) = \sigma_{1, t}^2$, Eq.~\ref{eq:Px} can be simplified to
\begin{equation}
 	P (x, x_0; t) = \frac{1}{\sqrt{2 \pi \sigma_{1, t}^2}} \exp \left[ - \frac{\left(x - x_0 e^{-t/\tau_k} \right)^2}{2 \sigma_{1, t}^2}\right].
 	\label{eq:Px1a}
\end{equation}%
The same equivalence relations hold for $y$ axis. Following the same steps, we obtain 
\begin{equation}
	P (y, y_0; t) = \frac{1}{\sqrt{2 \pi \sigma_{1, t}^2}} \exp \left[ - \frac{\left(y - y_0 e^{-t/\tau_k} \right)^2}{2 \sigma_{1, t}^2}\right].
	\label{eq:Py1a}
\end{equation}%
In the steady-state, i.e., at $t \gg \tau_k \gg \tau_{\mathrm{R}}$, $\sigma_{1, t}^2$ becomes $\sigma_1^2 = \frac{k_{\mathrm{B}} T}{k}\left(1 + \frac{\mathrm{Pe}^2}{2}\right)$, reducing Eqs.~\ref{eq:Px1a} and~\ref{eq:Py1a} to
\begin{align}
	P(x) & = \sqrt{\frac{k}{2 \pi k_{\mathrm{B}} T \left(1 + \frac{\mathrm{Pe}^2}{2}\right) }} \exp \left[ - \frac{k x^2}{2 k_{\mathrm{B}} T \left(1 + \frac{\mathrm{Pe}^2}{2}\right) }\right],\nonumber \\
	P(y) & = \sqrt{\frac{k}{2 \pi k_{\mathrm{B}} T \left(1 + \frac{\mathrm{Pe}^2}{2}\right)}} \exp \left[ - \frac{k y^2}{2 k_{\mathrm{B}} T \left(1 + \frac{\mathrm{Pe}^2}{2}\right) }\right].
	\label{eq:PxPy1SS}
\end{align}%
Therefore, the steady-state 2D position distribution is given by,
\begin{align}
	P(r) & = P(x) P(y) \\
	     & = \frac{k}{2 \pi k_{\mathrm{B}} T \left(1 + \frac{\mathrm{Pe}^2}{2}\right)} \exp \left[ - \frac{k r^2}{2 k_{\mathrm{B}} T \left(1 + \frac{\mathrm{Pe}^2}{2}\right)}\right],
	\label{eq:Pr1SSa}
\end{align}%
where $r^2=x^2+y^2$.
The steady-state position distributions, $P(x)$, $P(y)$, and $P(r)$ (Eq.~\ref{eq:PxPy1SS}, \ref{eq:Pr1SSa}) are zero-mean Gaussians with a standard deviation $\sigma_1 = \sqrt {\frac{ k_{\mathrm{B}} T}{k}\left(1 + \frac{\mathrm{Pe}^2}{2}\right)} $, which increases monotonically with activity, ${\mathrm{Pe}}$.
\vspace{2em}

\noindent
\textbf{Case-2: $\boldsymbol{\tau_{\mathrm{R}} \gg \tau_k}$} \\

\noindent This condition represents cases in which the harmonic confinement is very strong or the ABP has a very slow orientational diffusion. At this limit, the particle does not undergo any orientational diffusion, and the propulsion maintains its initial (at $t = 0$) direction for an extremely long time compared to its equilibration in the harmonic well. The initial propulsion direction is considered to be along the $x$-axis ($\phi_0$ = 0) without any loss of generality, because the harmonic potential is circularly symmetric. Therefore, for all practical purposes, $\tau_{\mathrm{R}}$ can be considered infinitely long, i.e., $\tau_{\mathrm{R}} \to \infty$; therefore, $\phi (t) = \phi_0$ = 0 for $t \lesssim \tau_{\mathrm{R}}$. \\

\noindent Consequently, Eq.~\ref{eq:x} can be written as,
\begin{align}
		            & x(t) - x_0 e^{-t/\tau_k} =  e^{-t/\tau_k} \int_{0}^{t} e^{t^{\prime}/\tau_k} \left[ v^x_{\mathrm{B}}(t^{\prime})+V \right] dt^{\prime} \\
		   \implies & x(t) - x_0 e^{-t/\tau_k} - V \tau_k \left( 1 - e^{-t/\tau_k}\right) \nonumber \\
		            & = e^{-t/\tau_k}\int_{0}^{t} e^{t^{\prime}/\tau_k} v^x_{\mathrm{B}}(t^{\prime}) dt^{\prime} \\
		   	        & = e^{-t/\tau_k} \int_{0}^{t} e^{t^{\prime}/\tau_k} A^x (t^{\prime})dt^{\prime}, \label{eq:xt2}
\end{align}
where $A^x(t^{\prime}) = v^x_{\mathrm{B}}(t^{\prime})$, which satisfies
\begin{equation}
	\left\langle A^x (t) \right\rangle = 0, \quad \text{and}  \quad  \left\langle A^x(t_1)A^x(t_2) \right\rangle = 2 D_{\mathrm{T}} \delta(t_1 - t_2).
	\label{AxProp2}
\end{equation}
This allows us to use the lemma in Eq.~\ref{eq:xt2}, where $\left(x - x_0 e^{-t/\tau_k} - V \tau_k \left( 1 - e^{-t/\tau_k}\right) \right)$, $e^{(t^{\prime} - t) / \tau_k}$, and $A^x (t)$ correspond to $R (t)$, $\psi(t^{\prime})$, and $A_{\mathrm{B}}(t)$, respectively. Thus, using the following equivalences 
\begin{align}
	q & \equiv D_{\mathrm{T}}, \quad \text{and} \nonumber \\ 
	\int_{0}^{t}\psi^2 (t^{\prime}) dt^{\prime} & \equiv  \int_{0}^{t} e^{(t^{\prime} - t) / \tau_k} dt^{\prime} = \frac{\tau_k}{2} \left( 1 - e^{- 2t / \tau_k} \right),
	\label{eq:equiv2}
\end{align}%
we obtain
\begin{equation}
	\begin{split}
		& P (x, x_0; t) \equiv P\left(x - x_0 e^{-t/\tau_k} - V \tau_k \left( 1 - e^{-t/\tau_k}\right) \right) \\ 
		& = \frac{1}{\sqrt{2 \pi D_{\mathrm{T}} \tau_k \left( 1 - e^{- 2t / \tau_k} \right)}} \times \\
		& \qquad \exp \left[ - \frac{\left(x - x_0 e^{-t/\tau_k} - V \tau_k \left( 1 - e^{-t/\tau_k}\right) \right)^2}{2 D_{\mathrm{T}} \tau_k \left( 1 - e^{- 2t / \tau_k} \right)}\right]. 
	\label{eq:Px2a}
	\end{split}
\end{equation}
\noindent
Similarly, when $\sin(\phi(t))$ = 0, $y$ position becomes
\begin{align}
	y(t) - y_0 e^{-t/\tau_k} & = e^{-t/\tau_k} \int_{0}^{t} e^{t^{\prime}/\tau_k} v^y_{\mathrm{B}}(t^{\prime}) dt^{\prime}  \nonumber \\ 
	                         & = e^{-t/\tau_k} \int_{0}^{t} e^{t^{\prime}/\tau_k} A^y (t^{\prime})dt^{\prime}, 
	\label{eq:yt2}
\end{align}
where $A^y(t^{\prime}) = v^y_{\mathrm{B}}(t^{\prime})$, which satisfies
\begin{equation}
	\left\langle A^y (t) \right\rangle = 0, \quad \text{and}  \quad  \left\langle A^y(t_1)A^y(t_2) \right\rangle = 2 D_{\mathrm{T}} \delta(t_1 - t_2).
	\label{AyProp2}
\end{equation}
Drawing parallel to the lemma, where $\left(y - y_0 e^{-t/\tau_k} \right)$, $e^{(t^{\prime} - t) / \tau_k}$, and $A^y (t)$ correspond to $R (t)$, $\psi(t^{\prime})$, and $A_{\mathrm{B}}(t)$, respectively, the equivalence relations in Eq.~\ref{eq:equiv2} hold for Eq.~\ref{eq:yt2}, we obtain 
\begin{align}
	& P (y, y_0; t) \equiv P\left(y - y_0 e^{-t/\tau_k} \right) \nonumber \\ 
	& = \frac{1}{\sqrt{2 \pi D_{\mathrm{T}} \tau_k \left( 1 - e^{- 2t / \tau_k} \right)}} \exp \left[ - \frac{\left(y - y_0 e^{-t/\tau_k} \right)^2}{2 D_{\mathrm{T}} \tau_k \left( 1 - e^{- 2t / \tau_k} \right)}\right]. 
	\label{eq:Py2a}
\end{align}%
This position distribution is the same as that of an HBBP, because the propulsion component along the $y$-direction continues to remain zero at $t < \tau_{\mathrm{R}}$.

\noindent
At $t \gg \tau_k$ (and $t \ll \tau_{\mathrm{R}}$), as the ABP equilibrates in the harmonic well, the position distributions (Eq.~\ref{eq:Px2a} and \ref{eq:Py2a}) are reduced to 
\begin{align}
	& P(x) = \sqrt{\frac{k}{2 \pi k_{\mathrm{B}} T}} \exp \left[ - \frac{k \left( x - V \tau_k \right) ^2}{2 k_{\mathrm{B}} T}\right], \quad \text{and} \nonumber \\
	& P(y) = \sqrt{\frac{k}{2 \pi k_{\mathrm{B}} T}} \exp \left[ - \frac{k y^2}{2 k_{\mathrm{B}} T}\right].
	\label{eq:PxPy2SS}
\end{align}
Thus, the 2D position distribution is given by
\begin{equation}
	P(x) P(y) = \frac{k}{2 \pi k_{\mathrm{B}} T} \exp \left[ - \frac{k \left( \left( x - V \tau_k \right) ^2 + y^2\right) }{2 k_{\mathrm{B}} T}\right] 
	\label{eq:Pxy2SS}
\end{equation}
This represents the position distribution of an HBBP at $x = V \tau_k$. At $ t > \tau_{\mathrm{R}}$, the slow reorientation of the propulsion direction results in a small nonzero tangential component, which leads to azimuthal motion of the ABP, whereas the substantial radial component of the propulsion continues to balance the restoring force at $r = V^r \tau_k$. Eventually, the bound HBBP trajectory moves freely along the periphery of a circle with radius $r = V \tau_k$, as the propulsion mostly points radially outward. Hence, at $t \gg \tau_{\mathrm{R}} \gg \tau_k$, the 2D steady-state position distribution takes the form,   
\begin{equation}
	P(r) = \frac{k}{2 \pi k_{\mathrm{B}} T} \exp \left[ - \frac{k  \left( r - V \tau_k \right) ^2 }{2 k_{\mathrm{B}} T}\right].
\end{equation}
The annular-shaped 2D position distribution appears bimodal when projected onto a single axis, where the peaks are located at $V \tau_k$, which increases linearly with the propulsion speed $V$.

\section{MSD calculations}\label{appendix:c}

The mean square displacement (MSD) along the $x$-direction is defined as 
\begin{align}
		\langle \Delta x^2 (t) \rangle  & = \langle (x(t) - x(0))^2\rangle \nonumber \\
										& = \langle x^2(t) \rangle + \langle x^2(0) \rangle - 2\langle x(t) x(0) \rangle,
		\label{eq:MSDx}
\end{align}%
where $x (t)$ and $x(0) \equiv x_0$ are given by Eq.~\ref{eq:x} and Eq.~\ref{eq:x0}, respectively, and $\left\langle ... \right\rangle $ denotes the ensemble average.\\
Therefore, following Eq.~\ref{eq:MSDx}, \ref{eq:MSDx1}, and \ref{eq:MSDx2}, the MSD along $x$-direction is given by, 
\begin{equation}
	\begin{split}
		\langle \Delta x^2 (t) \rangle & = 2 \left[  \langle x^2(t) \rangle - \langle x(t) x(0) \rangle\right]  \\
									   & = \frac{2k_{\mathrm{B}} T}{k} \left(1- e^{- t/ \tau_k} \right) \\
									   & \quad + \frac{ V^2 \tau_{\mathrm{R}} \tau^2_k }{ \left( \tau_{\mathrm{R}} + \tau_k\right) } \left[ 1 - \frac{\tau_{\mathrm{R}} e^{- t/ \tau_{\mathrm{R}}} - \tau_k e^{- t/ \tau_k}}{ \left( \tau_{\mathrm{R}} - \tau_k\right) }\right]  \\
									   & \quad + 2 \left( I_{2} - I_{1} \right) \cos2\phi_0.       
		\label{eq:MSDxTot}                                                                    
	\end{split}
\end{equation}%
Similarly, the MSD along the $y$-direction is derived as
\begin{equation}
	\begin{split}
		\langle \Delta y^2(t)\rangle & = \langle (y(t) - y(0))^2\rangle   \\
									 & = \frac{2k_{\mathrm{B}} T}{k} \left(1- e^{- t/ \tau_k} \right) \\
									 & \quad + \frac{ V^2 \tau_{\mathrm{R}} \tau^2_k }{ \left( \tau_{\mathrm{R}} + \tau_k\right) } \left[ 1 - \frac{\tau_{\mathrm{R}} e^{- t/ \tau_{\mathrm{R}}} - \tau_k e^{- t/ \tau_k}}{ \left( \tau_{\mathrm{R}} - \tau_k\right) }\right]  \\
									 & \quad - 2 \left( I_{2} - I_{1} \right) \cos2\phi_0,
		\label{eq:MSDy}
	\end{split}
\end{equation}%
Finally, the 2D MSD is obtained as 
\begin{equation}
	\begin{split}
		\langle \Delta r^2(t)\rangle & = \langle (r(t) - r(0))^2\rangle \\
									 & =  \langle (x(t) - x(0))^2\rangle + \langle (y(t) - y(0))^2\rangle \\
		                             & = \frac{4k_{\mathrm{B}} T}{k} \left(1- e^{- t/ \tau_k} \right) \\ 
									 & \quad + \frac{ 2 V^2 \tau_{\mathrm{R}} \tau^2_k }{ \left( \tau_{\mathrm{R}} + \tau_k\right) } \left[ 1 - \frac{\tau_{\mathrm{R}} e^{- t/ \tau_{\mathrm{R}}} - \tau_k e^{- t/ \tau_k}}{ \left( \tau_{\mathrm{R}} - \tau_k\right) }\right].
		\label{eq:MSDr}
	\end{split}
\end{equation}
\noindent
For the special case, when $ \tau_{\mathrm{R}} = \tau_k = \tau_{\mathrm{eq}}$, the expression for the MSDs (Eq.~\ref{eq:MSDxTot}, \ref{eq:MSDy}, and \ref{eq:MSDr}) diverge. This case can be treated by applying the equality condition at an early step in the derivation or by taking the limit $ \tau_{\mathrm{R}} \to \tau_k = \tau_{\mathrm{eq}}$ to obtain
\begin{equation}
	\begin{split}
		\langle \Delta r^2(t)\rangle & = \frac{4k_{\mathrm{B}} T}{k}  \left(1- e^{- t/ \tau_{\mathrm{eq}}} \right) \\
									 & \quad + V^2 \tau^2_{\mathrm{eq}} \left[ 1- e^{- t/ \tau_{\mathrm{eq}}}\left(1 - \frac{t}{\tau_{\mathrm{eq}}} \right) \right].
	\end{split}
\end{equation}

\section{Supplementary Videos}

\noindent
\href{https://drive.google.com/file/d/1RNH7BM5JcLSk9Q3rXtX4Jz0FPxZIqfZb/view?usp=drive_link}{\textbf{Supp. Video 1}} -- 
\label{vid:1}
\vspace*{0.1cm}
\noindent
Numerically simulated dynamics of a harmonically bound active Brownian particle in two dimensions with propulsion speed $V$ = \SI{10}{\micro\meter\per\second}, persistence time $\tau_{\mathrm{R}}$ = \SI{1}{\s}, equilibration time $\tau_k$ = \SI{100}{\s}, and ratio of characteristic timescales $\tau_{\mathrm{R}}/\tau_k$ = 0.01 (Fig.~\ref{fig:PosDist1}(b)). A magnified view of the active Brownian particle is presented as a silica Janus colloid (green-gray) with Pt-coating (gray) on one hemisphere. The propulsion direction, from the coated (gray) to the uncoated (green) side, indicated by a black arrow, evolves with the orientational diffusion of the Janus particle, while the speed ($V$) remains constant. A radially symmetric harmonic confining potential centered at (0, 0) is represented by an orange color gradient. Starting from the center, the particle exhibits bound dynamics, where the instantaneous direction of the resultant or residual velocity $\boldsymbol{v_{\mathrm{res}}}$ (Eq.~\ref{eq:ResVel}) is shown by a red arrow. Simulation time-step = \SI{1}{\ms}. (playback speed: 1$\times$)\\
\\
\noindent
\href{https://drive.google.com/file/d/16bpfX74Cn7hS5IDzzbZ9Xkn4gPCN1FPX/view?usp=drive_link}{\textbf{Supp. Video 2}} -- 
\vspace*{0.1cm}
\noindent
Numerically simulated dynamics of a harmonically bound active Brownian particle in two dimensions with propulsion speed $V$ = \SI{10}{\micro\meter\per\second}, persistence time $\tau_{\mathrm{R}}$ = \SI{100}{\s}, equilibration time $\tau_k$ = \SI{1}{\s}, and ratio of the characteristic timescales $\tau_{\mathrm{R}}/\tau_k$ = 100 (Fig.~\ref{fig:PosDist2}(b)). The active Brownian particle is shown as a silica Janus microsphere (green-gray) with a Pt-coating (gray) on one hemisphere. While the propulsion speed ($V$) remains constant, its direction, from the coated (gray) to the uncoated (green) side, indicated by a black arrow, evolves with the orientational diffusion of the Janus particle. The orange gradient represents the radially symmetric harmonic confining potential centered at (0, 0). Starting from the center, the particle mostly follows the initial propulsion direction, which is along the $x$-axis, until the propulsion changes its direction with slow orientational diffusion of the particle. The instantaneous direction of the resultant or residual velocity $\boldsymbol{v_{\mathrm{res}}}$ (Eq.~\ref{eq:ResVel}) is indicated by a red arrow. The propulsion is counteracted by the restoring force (${-} k\boldsymbol{r}$) and eventually balanced at a radial distance $r = V \tau_k$ (marked by a dashed circle) (Eq.~\ref{eq:ResVel}). Thus, the particle remains confined along the radial direction at $r = V \tau_k $, while performing a free and slow dynamics along the azimuthal direction, as the propulsion mostly points radially outwards. Simulation time-step = \SI{1}{\ms}. (playback speed: 1$\times$)\\
\\
\noindent
\href{https://drive.google.com/file/d/1EToFYCh39pHFejzeOe8I-CffrSpCgkeZ/view?usp=drive_link}{\textbf{Supp. Video 3}} -- 
\vspace*{0.1cm}
\noindent
Numerically simulated dynamics of a harmonically bound active Brownian particle in two dimensions with propulsion speed $V$ = \SI{10}{\micro\meter\per\second}, and both the persistence time $\tau_{\mathrm{R}}$ and equilibration time $\tau_k$ = \SI{10}{\s}; therefore, the ratio of the characteristic timescales $\tau_{\mathrm{R}}/\tau_k$ = 1 (Fig.~\ref{fig:PosDist2}, green). A magnified view of the active Brownian particle is presented as a silica Janus colloid (green-gray) with Pt-coating (gray) on one hemisphere. The propulsion direction, from the coated (gray) to the uncoated (green) side, indicated by a black arrow, evolves with the orientational diffusion of the Janus particle, while the speed ($V$) remains constant. A radially symmetric harmonic confining potential centered at (0, 0) is indicated by an orange color gradient. Starting at the center, the particle exhibits bound dynamics, where the instantaneous direction of the resultant or residual velocity  $\boldsymbol{v_{\mathrm{res}}}$ (Eq.~\ref{eq:ResVel}) is shown by a red arrow. Simulation time-step = \SI{1}{\ms}. (playback speed: 1$\times$)

\bibliography{ABPinHW_references}

\end{document}